\documentclass{article}

\usepackage{xcolor,colortbl}
\usepackage{microtype}
\usepackage{graphicx}
\usepackage{subfigure}
\usepackage{multirow}
\usepackage{amsmath}
\usepackage{booktabs} 
\usepackage{tablefootnote}
\usepackage[hyphens]{url}
\usepackage{hyperref}
\usepackage{caption}

\usepackage{xspace}
\usepackage{xargs} 
\usepackage[colorinlistoftodos,prependcaption,textsize=tiny]{todonotes}
\newcommandx{\richard}[2][1=]{\todo[linecolor=blue,backgroundcolor=blue!20,bordercolor=blue,#1,size=tiny]{RC: #2}\xspace}
\usepackage{xspace}
\usepackage{xcolor}
\usepackage{CJKutf8}
\usepackage{pifont}
\usepackage{tikz}

\usepackage[accepted]{icml2022}

\icmltitlerunning{Alignment-Aware Acoustic \& Text Pretraining for 
Speech Synthesis \& Editing}

\usepackage{dsfont}
\usepackage{amsmath}
\usepackage{amssymb}
\usepackage{amsthm}
\usepackage{bm} 
\usepackage{balance}

\usepackage[utf8]{inputenc}
\newcolumntype{N}{@{}m{0pt}@{}}

\newcommand{\tuple}[1]{\ensuremath{\langle {#1} \rangle}}

\newcommand{\notes}[1]{}

\theoremstyle{definition}

\theoremstyle{plain}

\newcommand{\vecs}{\mathbf{s}\xspace}

\newcommand{\aaat}{A$^3$T\xspace}

\newcommand{\vecd}{\ensuremath{\mathbf{d}}}

\newcommand{\ith}[1]{\ensuremath{i^{{th}}}}

\newcount\permx
\newcount\permy
\def\permdot#1#2{
\permx=#1 \advance\permx by-1
\permy=#2 \advance\permy by-1
\psframe[fillcolor=black, fillstyle=solid]
(\permx,\permy)(#1, #2)
}

\newcommand{\boxnum}[1]{{\setlength{\fboxsep}{1pt}\raisebox{1pt}{\hspace{1pt}\fbox{\tiny #1}\hspace{1pt}}}}
\newcommand{\ind}[1]{\ensuremath{_{\kern-0.5pt\boxnum{#1}}}}

\newcommand{\vecx}{\mathbf{x}\xspace}

\newcommand{\vece}{\ensuremath{\bm{e}}\xspace}

\newcommand{\smallnt}[1]{\ensuremath{_{\mbox{\tiny PP}}}\xspace}

\newcommand{\pseudocode}{Algorithm}
\floatname{algorithm}{\pseudocode}

\iffalse

\else

\fi

\newcommand{\dsx}{D_{\vecs,\vecx}\xspace}

\begin{document}

\begin{CJK}{UTF8}{gbsn}

\twocolumn[
\icmltitle{
\aaat: Alignment-Aware Acoustic and Text Pretraining for \\
Speech Synthesis and Editing
}



\icmlsetsymbol{equal}{*}

\begin{icmlauthorlist}
\icmlauthor{He Bai}{equal,waterloo}
\icmlauthor{Renjie Zheng}{equal,baidu}
\icmlauthor{Junkun Chen}{osu}
\icmlauthor{Xintong Li}{baidu}
\icmlauthor{Mingbo Ma}{baidu}
\icmlauthor{Liang Huang}{osu}
\end{icmlauthorlist}

\icmlaffiliation{waterloo}{University of Waterloo, Waterloo, ON, Canada (work done at Baidu Research USA)}
\icmlaffiliation{baidu}{Baidu Research, Sunnyvale, CA, USA}
\icmlaffiliation{osu}{Oregon State University, Corvallis, OR, USA}

\icmlcorrespondingauthor{Renjie Zheng}{zrenj11@gmail.com}
\icmlcorrespondingauthor{Liang Huang}{liang.huang.sh@gmail.com}

\icmlkeywords{Machine Learning}

\vskip 0.3in
]



\printAffiliationsAndNotice{\icmlEqualContribution}  

\begin{abstract}
Recently, speech representation learning 
has  improved many speech-related tasks
such as speech recognition, speech classification,
and speech-to-text translation.
However, 
all the above tasks are in the direction of
{\em speech understanding}, 
but for the inverse direction,
{\em speech synthesis}, 
the potential of representation learning is yet to be realized, 
due to the challenging nature of generating
high-quality speech.
To address this problem, 
we propose our framework, 
Alignment-Aware Acoustic-Text Pretraining (\aaat),
which 
reconstructs masked acoustic signals with text input and acoustic-text alignment during training.
In this way, the pretrained model can generate high quality reconstructed spectrogram, which can be applied to the speech editing and unseen speaker TTS directly.
Experiments show \aaat outperforms
SOTA models on speech editing,
and improves
multi-speaker speech synthesis without the external speaker verification model.\footnote{See demos at {\scriptsize \url{https://educated-toothpaste-462.notion.site/Demo-b0edd300e6004c508744c6259369a468}}. Code available at: {\scriptsize \url{https://github.com/richardbaihe/a3t}}}
\end{abstract}

\section{Introduction}


Recently, 
speech representation learning
has attracted much attention in the speech community due to its
strong performance to many
speech-related downstream tasks,
such as speech recognition, speech classification, and speech translation
\cite{baevski2020wav2vec,chen2020mam, liu2020tera, zheng2021fused, hsu2021hubert}

However, all these efforts can only support 
{\em speech understanding} tasks 
which take speech as input,
but for the inverse direction,
{\em speech synthesis}, which synthesis speech as output,
the potential of representation learning is yet to be realized. 
For example, one line of work,
such as wav2vec 2.0 \cite{baevski2020wav2vec},
Hubert \cite{hsu2021hubert} and SLAM \cite{bapna2021slam},
learn discrete quantized speech units 
as latent representations.
In this way, these models are good at recognizing and 
extracting discrete information from speech and successfully improves
automatic speech recognition (ASR), but they are unable to generate continuous
acoustic signals for speech synthesis.
On the other hand, another line of work,
such as MAM \cite{chen2020mam} and FAT-MLM \cite{zheng2021fused},
show that reconstructing masked spectrogram with continuous units can improve
speech-to-text translation.
However, the quality of their proposed speech reconstruction 
is far from the requirement of speech synthesis tasks 
(see Fig.~\ref{fig:ljs_ab_f}).

\begin{table}[t!]
	\centering
\resizebox{1.0\columnwidth}{!}{
		\begin{tabular}{l|c|c|c|c|c}
\toprule
      & \multicolumn{2}{c|}{\textbf{Data}} & \textbf{Reconstructed} & \multicolumn{2}{c}{\textbf{Tasks}} \\ 
\textbf{Model}  & \multirow{2}{*}{$\langle \vecs, \vecx \rangle$} &  \multirow{2}{*}{$\langle \vecs \rangle$}  & \textbf{Masked } & \textit{Speech } & \textit{Text-to-} \\ 
        &   &     & \textbf{ Speech} & \textit{Editing} & \textit{Speech} \\ 
\midrule
wav2vec 2.0            &                             & \checkmark                  &                  &                             \\
Hubert                 &                             & \checkmark                  & discrete units    &                             \\
SLAM                   & \checkmark                  & \checkmark                  &                  &                             \\
\midrule
MAM                    &                             & \checkmark                  & low-quality      &                             \\
FAT-MLM                & \checkmark                  & \checkmark                  & spectrogram      &                             \\
\midrule
\multirow{2}{*}{\aaat} & \multirow{2}{*}{\checkmark} & \multirow{2}{*}{\checkmark} & high-quality     & \multirow{2}{*}{\checkmark}  & \multirow{2}{*}{\checkmark} \\
                       &                             &                             & spectrogram      &                              &                             \\ 
\midrule
\end{tabular}
}
\caption{Comparisons of \aaat  
with other existing speech pretraining
models. Here $\vecs$ stands for speech input, while $\vecx$ stands for 
text, and $\tuple{\vecs, \vecx}$ denotes parallel speech-text data.}
\vspace{-0.5cm}
\label{table:intro}
\end{table}

\begin{figure*}[ht!]
\centering
\begin{tabular}{l}
\begin{minipage}[b]{1.0 \linewidth}
\begin{center}
\subfigure[Wav2vec 2.0.]{
\makebox[0.5\linewidth][c]{
\includegraphics[width=8cm]{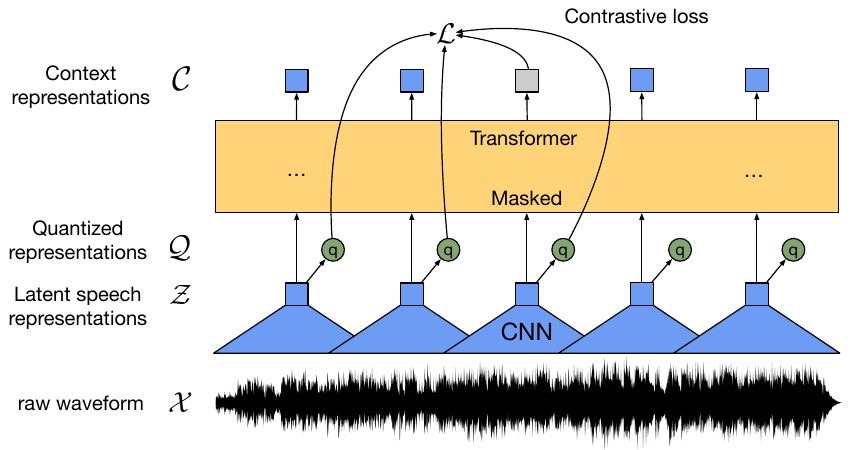}
    }
\label{fig:wav2vec2}
}
\subfigure[Masked Acoustic Model (MAM).]{
\includegraphics[width=6.cm]{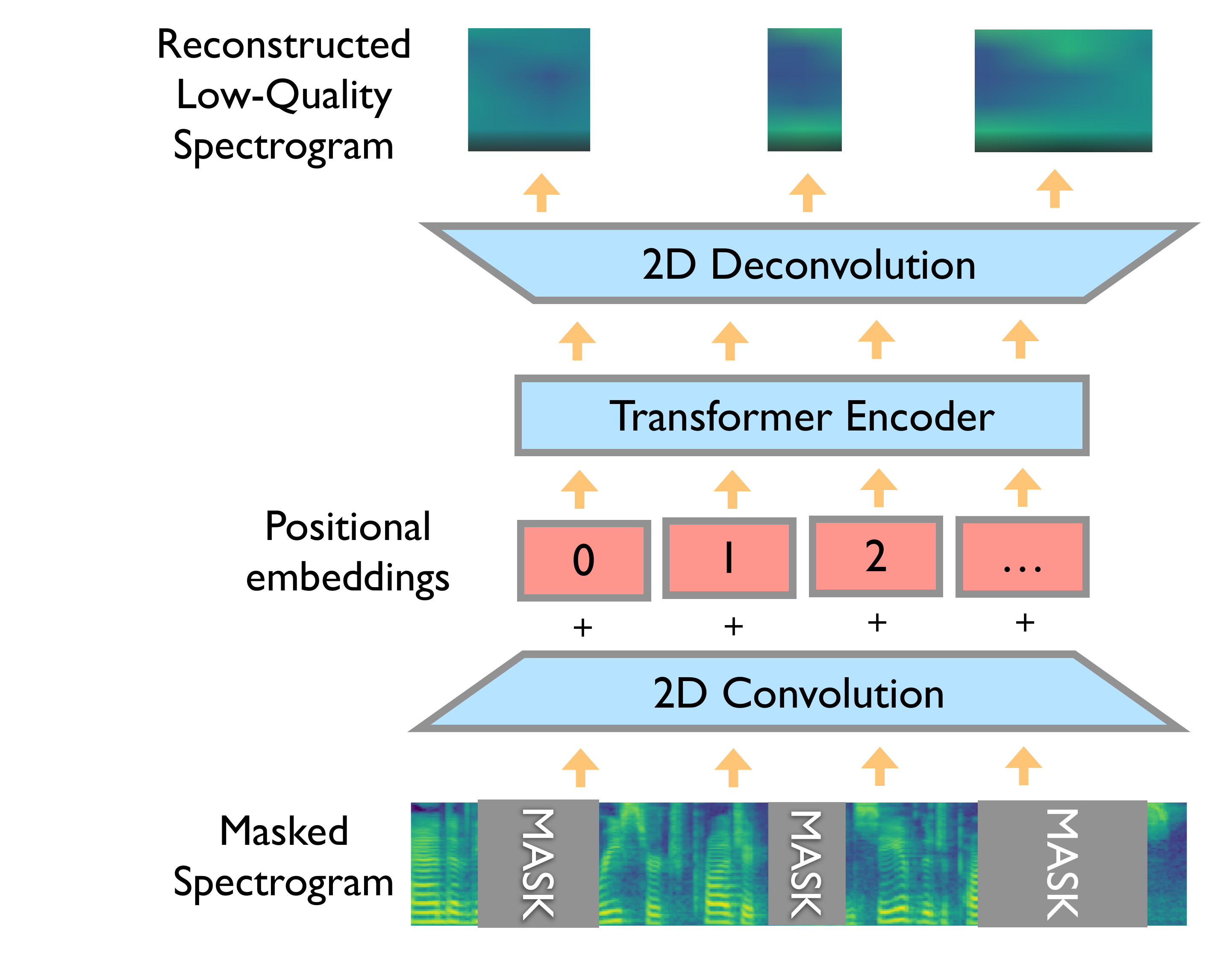}
\label{fig:mam}
}
\end{center}
\end{minipage}
\\[-0.1cm]
\begin{minipage}[b]{1.0 \linewidth}
\begin{center}
\subfigure[FAT-MLM.]{
\includegraphics[width=15cm]{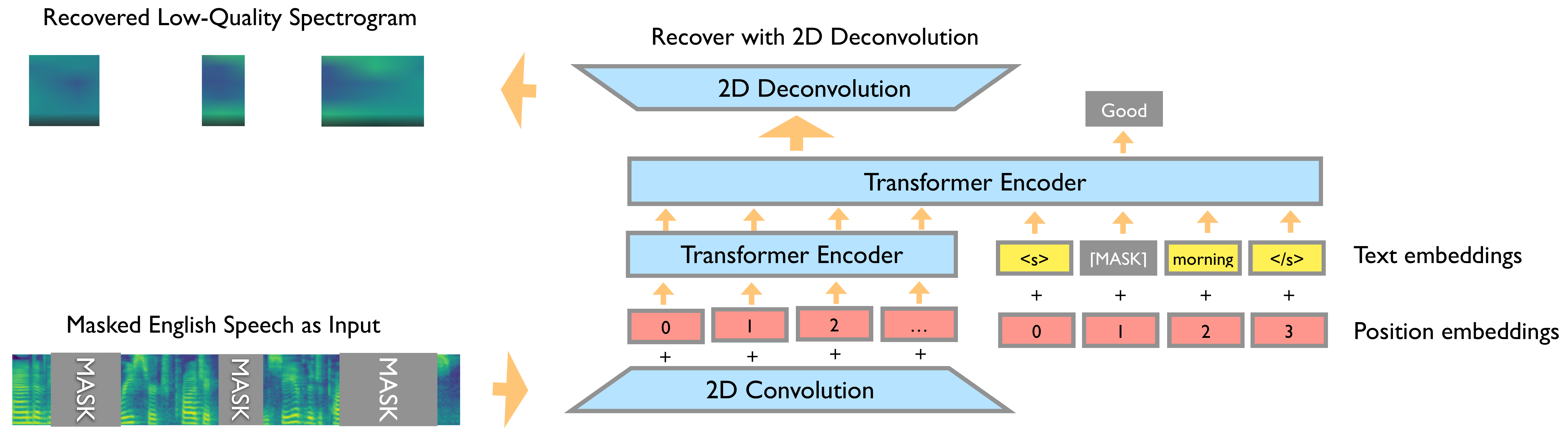}
\vspace{-0.9cm}
\label{fig:fat_mlm}
}
\end{center}
\end{minipage}
\end{tabular}
\vspace{-0.3cm}
\caption{Previous work for speech 
representation learning.
}
\vspace{-0.5cm}
\label{fig:exist_work}
\end{figure*}

To address this problem, we propose our framework, 
Alignment-Aware Acoustic-Text Pretraining 
(\aaat),
where we introduce cross-modal alignment embeddings
which make the model easier to learn the alignment between
the acoustic and phoneme 
input during multi-modal pretraining, 
and significantly improve the quality of the reconstructed acoustic signals.
Different from the segment embeddings used in Segatron and SegaBERT~\cite{bai2021segatron,bai-etal-2022-better}, which improve language modeling by grouping tokens according to the sentence position and paragraph position, our alignment embeddings can align a phoneme and its frames together to learn the cross-modal self-attention~(Fig.~\ref{fig:attention_heatmap}).
Moreover, we borrow several useful ideas from recent 
text-to-speech (TTS)
literature, including Conformer~\cite{gulati2020conformer, guo2021recent} and Post-Net~\cite{shen2018natural}, to further improve
the quality of our reconstructed spectrograms.

Without any finetuning, the proposed model can be adopted as a speech-editing system,
a task that modifies an existing speech,
by reconstructing the desired acoustic signals given
original contextual speech and modified text.
Furthermore, the model can be adopted as a multi-speaker TTS system with our proposed prompt-based decoding method, to synthesis unseen speaker's speech without the external speaker verification model~(speaker embeddings).
Our experiments show that our \aaat with prompt-based decoding can outperform the TTS model equipped with both the speaker embedding~\cite{jia2018transfer} and the global style token~(GST)~\cite{wang2018style}.

We make the following contributions:

\begin{itemize}
\vspace{-0.3cm}
    \item We propose the Alignment-Aware Acoustic-Text Pretraining 
    (\aaat), a BERT-style pretraining model,
    which takes both phonemes and partially-masked spectrograms as inputs. 
    It can reconstruct masked spectrograms with high quality without finetuning and uses the identical framework for decoding.

    \item We show that the proposed \aaat model has the ability
    to do speech editing and outperforms the current SOTA.

    \item We propose the prompt-based decoding method. We show that our \aaat model has the ability to do speech synthesis for unseen speaker and outperforms the speaker-embedding-based multi-speaker TTS system.
    
\end{itemize}

\section{Previous Work}



\subsection{Speech Synthesis and Editing}

Recently, neural TTS systems become capable of generating audios with high naturalness~\cite{oord+:2016,shen+:2018,ren+:2019,peng+:2019, ren2020fastspeech}.
SOTA neural TTS systems generally 
consist of two stages: 
the {\em text-to-spectrogram} stage which generates
an intermediate acoustic representation (linear- or mel-spectrogram) from the text,
and the {\em spectrogram-to-wave} stage~(vocoder)
which converts the aforementioned acoustic representation into actual wave signals~\cite{oord+:2018,prenger+:19}.\footnote{We focus on the {\em text-to-spectrogram} 
stage and use an off-the-shelf vocoder Parallel WaveGAN~\cite{yamamoto2020parallel}.}

In the multi-speaker and unseen-speaker settings, the existing TTS models need to be trained with an additional input feature: speaker embedding~\cite{jia2018transfer}, which is extracted from an external speaker verification model trained with tens of thousands of speakers' audio.
And during the inference for an unseen speaker, the embedding will be extracted from one of this speaker's other audio examples.
However, the embedding from the speaker verification model is not optimized directly to capture speaker characteristics relevant to synthesis, and cannot provide enough information for the TTS model to generate audio similar to the example.

The input of speech editing 
includes the original speech, the original and modified text.
\citet{jin2017voco} propose to insert 
a regenerated audio clip back into the original recording.
However, due to the absence of speech contextual information, the boundaries of the modified region would be not smooth.
\citet{morrison2021context} propose to retrieve the modified speech segments from other utterances of the same speaker and correct the prosody with a context-aware TD-PSOLA corrector~\cite{moulines1990pitch}.
However, the edited content may not be found in the speech data of the same speaker.
Most recently,~\cite{tan2021editspeech} use neural TTS model to generate better-modified speech.
This method is only compatible with auto-regressive decoding models and highly relies on the speaker embeddings, which limits its efficiency and transferability to new speakers.

\subsection{Speech Pretraining}

To improve the Text-to-Speech model from larger-scale pure speech data,
one idea is to do pretraining on speech data.
All existing speech pretraining work learn either 
discrete units, which can only support speech understanding tasks, 
or spectorgram, but with very low quality.

\subsubsection{Reconstructing Discrete Units}

Wav2vec 2.0 proposed by~\citet{baevski2020wav2vec}
is the most popular speech pretrain model recently.
It masks the speech input 
in the latent space and pretrains the model 
by predicting discrete units
via a contrastive task defined
over a quantization of the latent representations,
as shown in Fig.~\ref{fig:wav2vec2}.
Similar to wav2vec 2.0, Hubert~\cite{hsu2021hubert} 
and SLAM~\cite{bapna2021slam} also learn
discrete speech units from contextualized representations to 
represent the latent representations. 
Thus these models can achieve good performance in
speech recognition tasks, but they are
unable to generate continuous acoustic signals for speech
synthesis. 

\subsubsection{Reconstructing Low-Quality Spectrogram}

Recently,~\citet{chen2020mam}
propose to learn a speech encoder in a self-supervised fashion 
on the speech side, which can utilize speech data without transcription.
Fig.~\ref{fig:mam} demonstrate the architecture of this model, termed Masked Acoustic Modeling (MAM). 
MAM replaces a span of speech spectrogram with mask tokens,
and learns to recover
the masked spectrogram during training.
On the other hand, 
\citet{zheng2021fused} propose a Fused Acoustic and Text Masked Language Model (FAT-MLM) which jointly learns a unified representation for both acoustic and text input from various types of corpora including parallel data for speech recognition and machine translation, and even pure speech and text data,
as shown in Fig.~\ref{fig:fat_mlm}. 

Both MAM and FAT-MLM reconstruct 
spectrograms,
however, the quality of their spectrogram output
is far from the requirement of speech synthesis tasks (see Fig.~\ref{fig:ljs_ab_f}),
since these pretrained models are all used in 
speech understanding task (speech-to-text translation),
where the quality of the reconstructed spectrogram is not 
very important.

\begin{figure}[ht!]
\centering
\begin{tabular}{c}

\begin{minipage}[b]{1.0 \linewidth}
    \begin{center}
    \subfigure[Forced Alignment Preprocessing.]{
        \includegraphics[width=6.0cm]{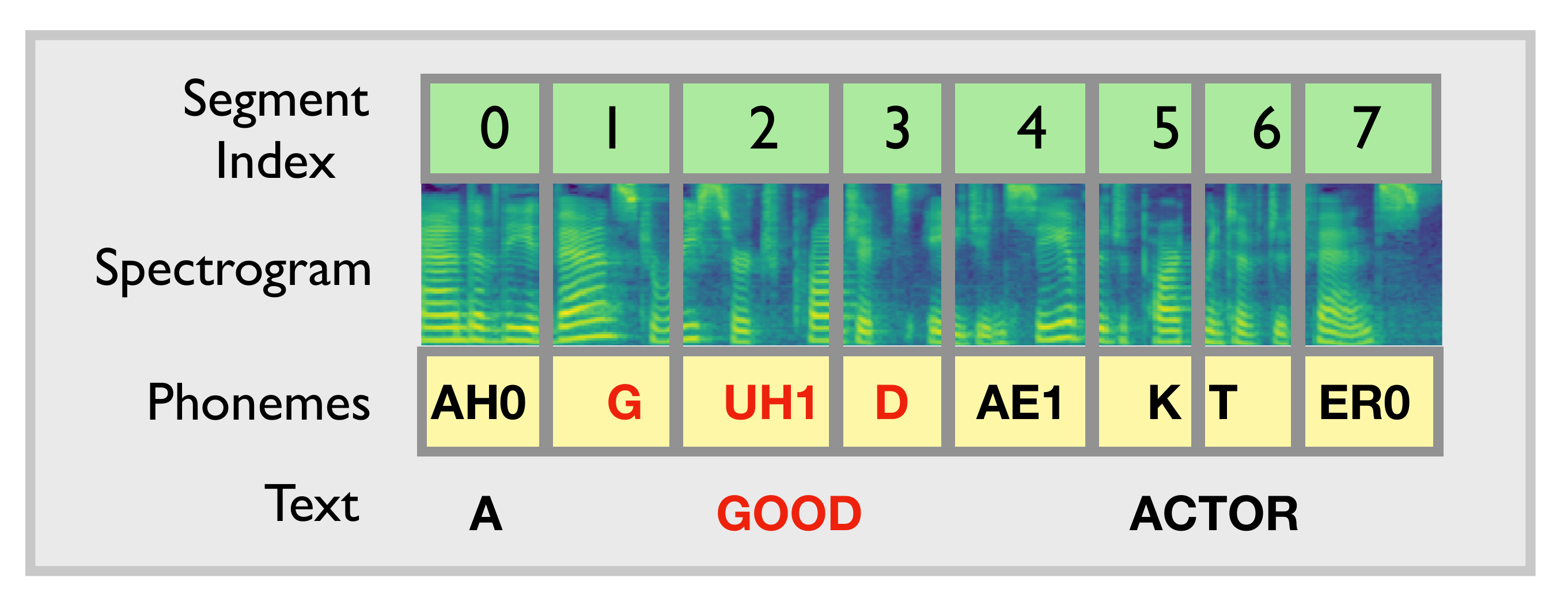}
        \label{fig: forced_alignment}
    }
    \end{center}
    \end{minipage}
    \\

\begin{minipage}[b]{1.0 \linewidth}
\begin{center}
\subfigure[A$^3$T Model. Segment embeddings are assigned according to the Segment Index produced by forced alignment preprocssing (see  subfigure (a)).]{
    \includegraphics[width=8.0cm]{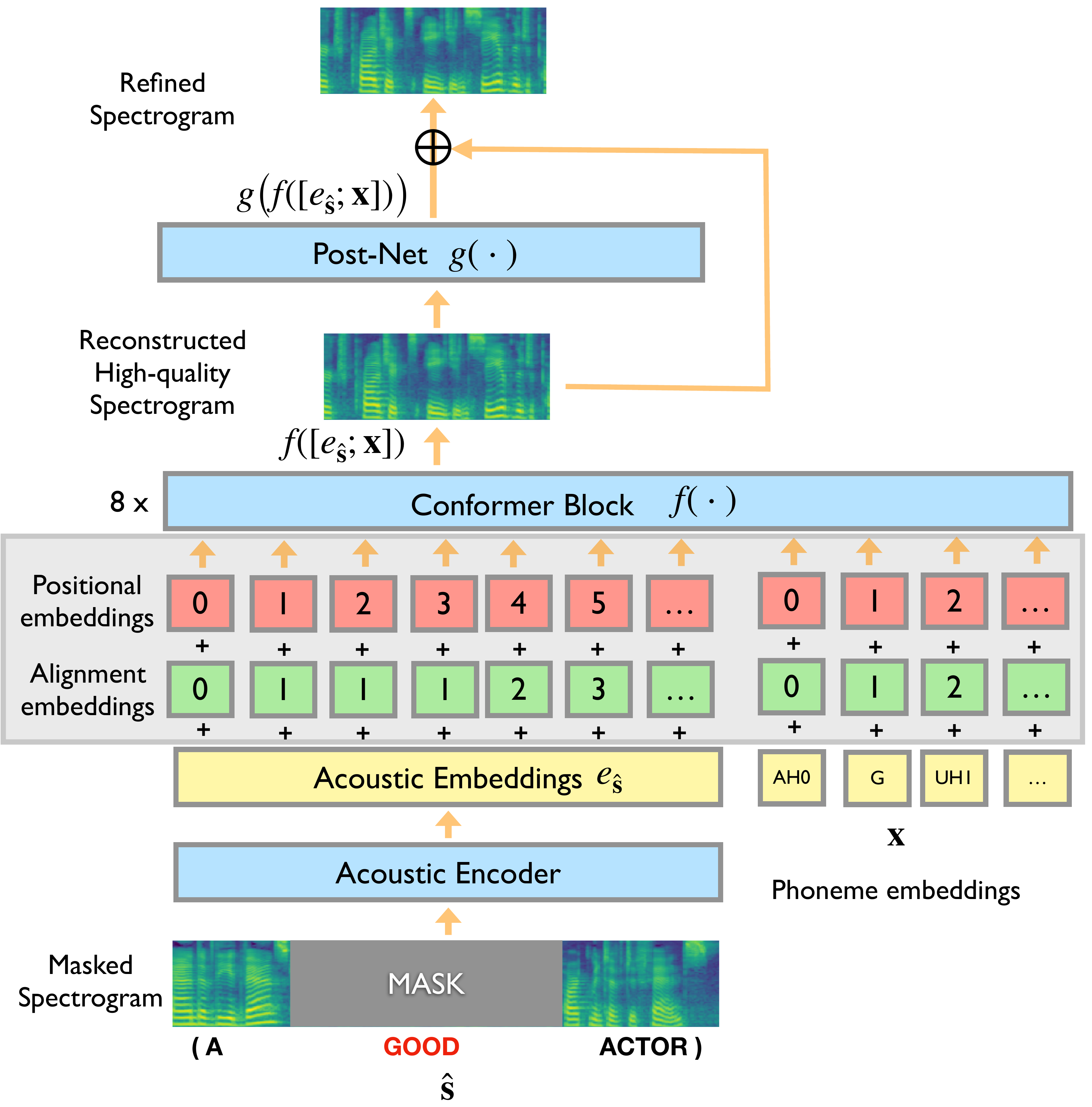}
}
\end{center}
\end{minipage}
\\
\begin{minipage}[b]{0.5 \linewidth}
    \begin{center}
    \subfigure[Conformer Block.]{
        \includegraphics[width=3.7cm]{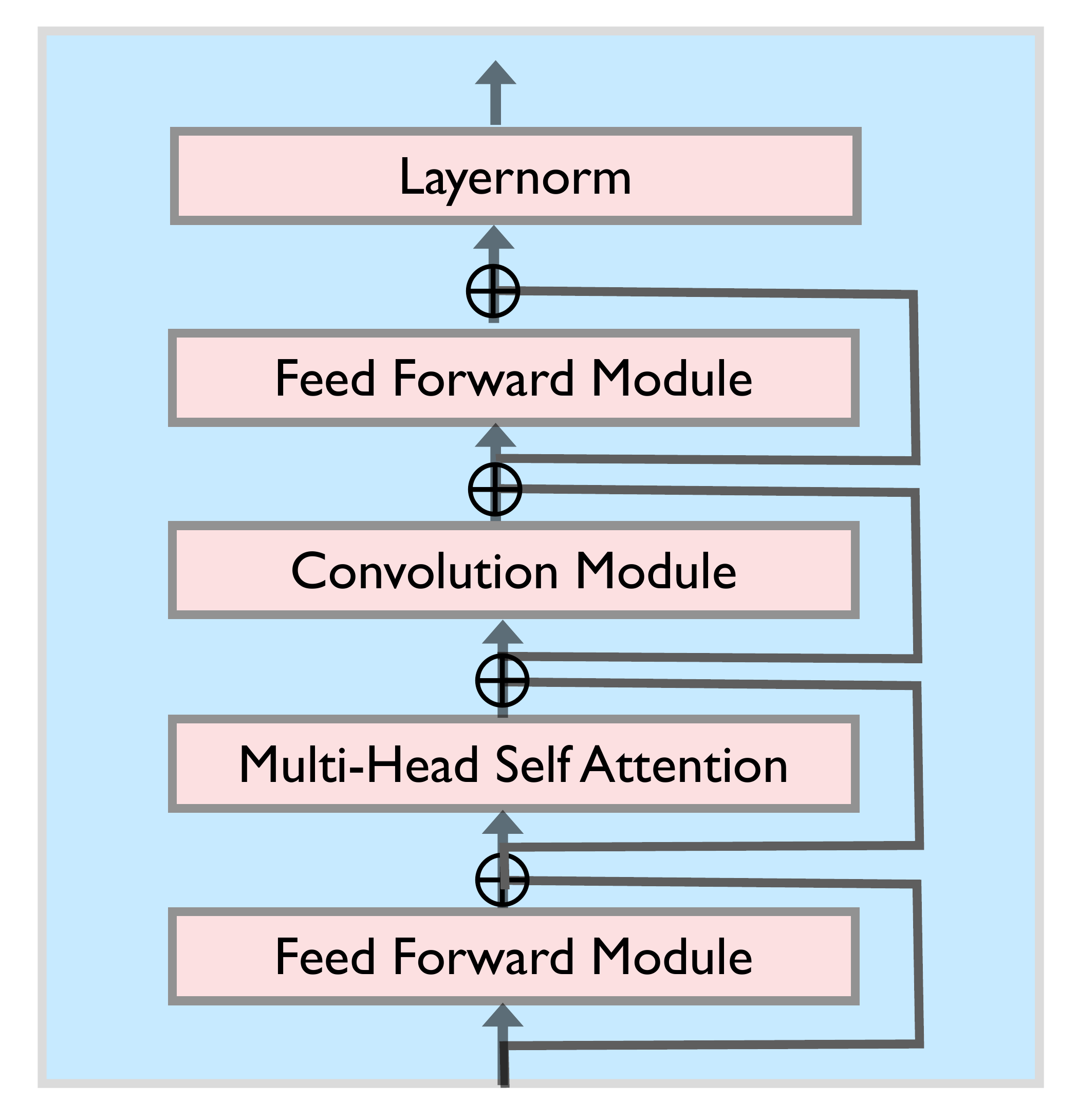}
        \label{fig: conformer}
    }
    \end{center}
\end{minipage}

\begin{minipage}[b]{0.5 \linewidth}
    \begin{center}
    \subfigure[Post-Net.]{
        \includegraphics[width=3.7cm]{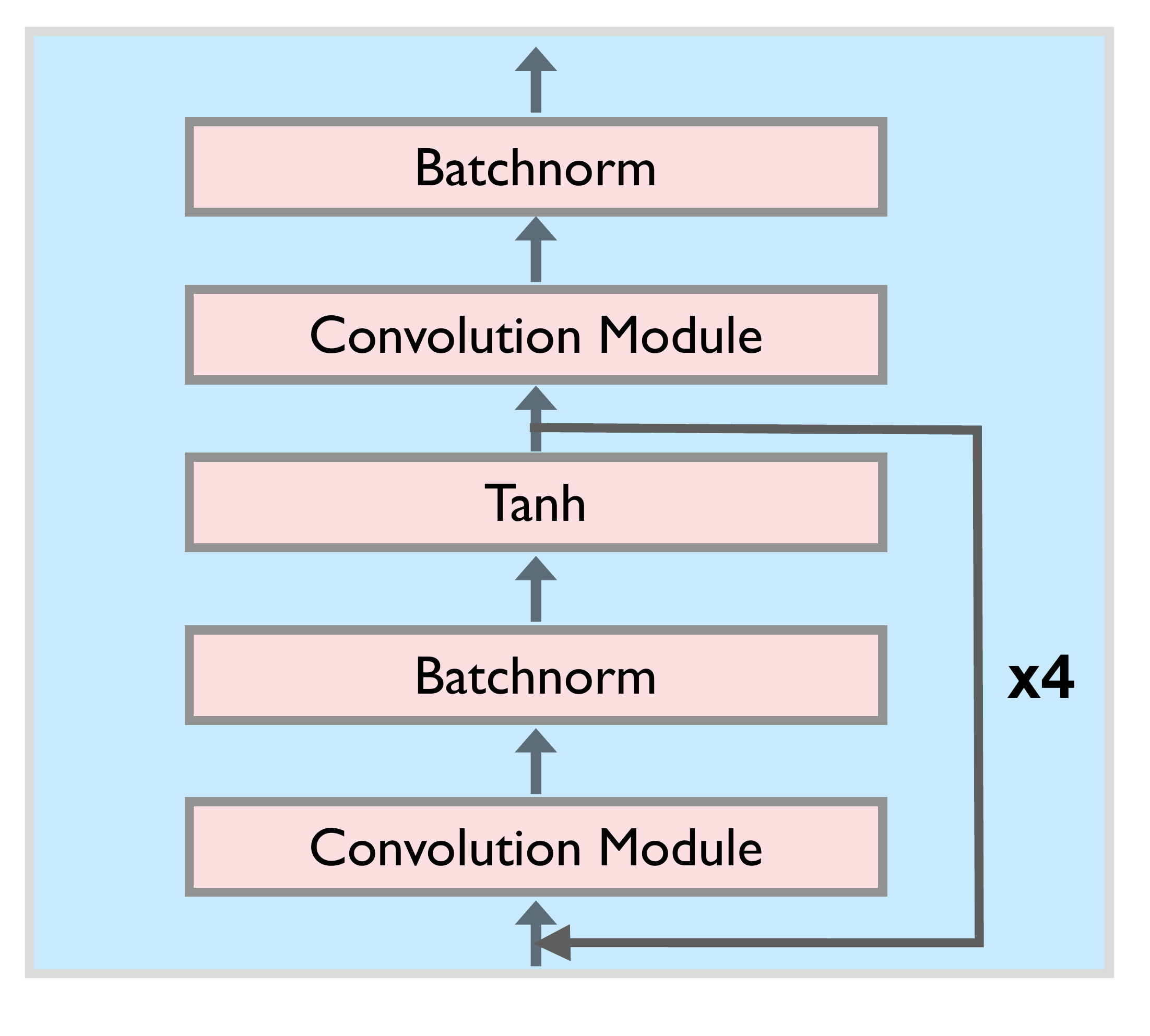}\label{fig: postnet}
    }
    \end{center}
\end{minipage}

\end{tabular}
\caption{Alignment-Aware Acoustic-Text Pretraining 
(A$^3$T).}
\vspace{-0.5cm}
\label{fig:fat-tts}
\end{figure}

\section{Alignment-Aware Acoustic-Text Pretraining }


Although existing speech pretraining models show a strong 
representation learning ability and significantly improve upon many down-stream tasks in {\em speech understanding}, 
all these efforts can not support 
{\em speech synthesis} tasks.
To address this problem, we propose the 
Alignment-Aware Acoustic-Text Pretraining (\aaat)
which learns to generate high-quality spectrogram
given speech context and text.

\subsection{\aaat}

\aaat takes speech and
transcription tuples as input, denotes as $D_{\vecs, \vecx} = \{\langle \vecs, \vecx \rangle^{(n)} \}^{|D|}_{n=1}$, 
where $\vecs=(s_1, ... , s_{|s|})$ is a sequence of 
acoustic features 
 $s_i \in \mathbb{R}^{d_\text{s}}$ 
which 
can be the spectrogram 
or mel-spectrogram of the 
speech audio,
and each $s_i$ represents the frame-level speech 
feature, 
and $\vecx=(x_1, ... , x_{|\vecx|})$ is the sequence of corresponding transcription.

As shown in Fig.~\ref{fig:fat-tts}, 
we first
randomly mask several spans of $\vecs$
by a random masking function over the 
input $\vecs$:
$\hat{\vecs} \sim  \text{Mask}_{\text{span}}(\vecs, \lambda)$,
where $\text{Mask}_{\text{span}}(\cdot)$ 
replaces several random spans of $\vecs$
by the probability of $\lambda$
with the same number of a random initialized masking vector $\epsilon_{\vecs} \in \mathbb{R}^{d_{\text s} }$.
Then we encode $\hat{\vecs}$ with a acoustic encoder
for acoustic embeddings $\vece_{\hat{\vecs}}$.
In this work, we use a nonlinear feed-forward layer as the acoustic encoder.

\subsection{Cross-modal Alignment Embedding}
To strengthen the interaction between the speech and text input, we introduce cross-modal alignment embedding as one input of encoder, where we sum the $i$th acoustic embedding $\vece_{s_i}$ or text embedding $\vecx_i$ with its positional embedding $\vece_{\text{pos}_i}$ and alignment
embedding $\vece_{\text{aln}_i}$ all together: $\vece_{s_i}+\vece_{\text{pos}_i}+\vece_{\text{aln}_i}$,
where previous work have proved the embedding sum operation is simple and effective~\cite{devlin2018bert, bai2021segatron}.
After that, the phoneme embedding and its acoustic embeddings will share the same alignment embedding. 
We use a forced aligner ~\cite{yuan2008speaker} to pre-process the dataset to get the alignment information, which is shown in Fig.~\ref{fig: forced_alignment}.

\subsection{Conformer}
Given the recent success of Convolution-augmented Transformer~(Conformer) on various speech tasks~\cite{gulati2020conformer, guo2021recent}, we adopt Conformer as the backbone of our encoder and decoder.
Compared with Transformer, Conformer introduces a convolution module and an additional feedforward module, which is shown in Fig.~\ref{fig: conformer}.
In our experiments, we find Conformer is better than Transformer for acoustic-text pretraining.

\subsection{Post-Net and Loss Function}

We follow Tacotron 2 \cite{shen2018natural} to use Post-Net to refine the generated spectrogram.
The predicted spectorgram is passed through a 5-layer convolution Post-Net to be refined as shown in Fig.~\ref{fig: postnet}.

The training objective of multi-modal \aaat
includes a speech reconstruction loss $\ell_{\vecs}(\dsx)$
which takes a spectrogram $\vecs$ and a text sequence $\vecx$ 
as input.
We have the following training objective to reconstruct the 
original speech signal with the surrounding context 
information:\footnote{Similar with previous work 
using masked
language model objective,
this loss only takes the masked input into consideration.}
\begin{equation}
\begin{aligned}
\ell_{\vecs}(D_{\vecs, \vecx}) = \displaystyle\sum_{\langle\vecs, \vecx\rangle \in D_{\vecs, \vecx}}  &\| \underbrace{f ([e_{\hat{\vecs}}; \vecx]) + g \big( f ([e_{\hat{\vecs}}; \vecx])\big)}_{\text{refined spectrogram}} - \vecs  \|_1
\\
&+\| \!\!\!\!\!\!\!\!\!\! \underbrace{f ([e_{\hat{\vecs}}; \vecx])}_{\text{reconstructed spectrogram} } \!\!\!\!\!\!\!\!\!\! - \vecs  \|_1
\label{eq:reconstruct}
\end{aligned}
\end{equation}
where $g$ is a Post-Net  
which tries to recover a better original
signal from encoded representation $f([e_{\hat{\vecs}}; \hat{\vecx}])$.
We use
mean absolute error~(MAE) for measuring the difference
between $s$ and the reconstructed spectrogram.


\begin{figure}[t!]
    \centering
    \begin{tabular}{c}
    \\
    \begin{minipage}[b]{1.0 \linewidth}
    \begin{center}
    \includegraphics[width=7.cm]{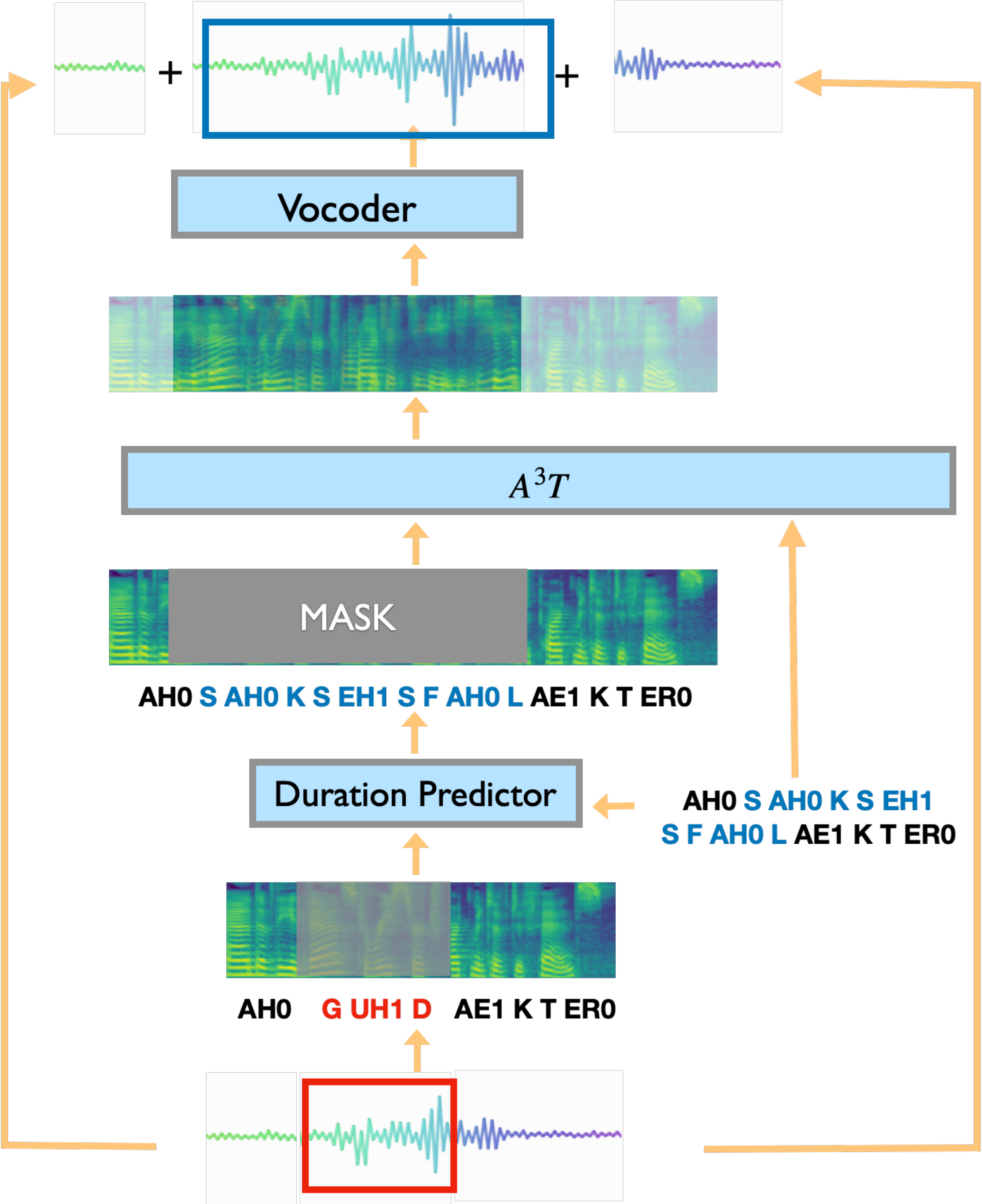}
    \end{center}
    \end{minipage}
    \end{tabular}
    \caption{Speech editing pipeline.}
    \label{fig:speech_edit}
\end{figure}

\subsection{\aaat for Speech Editing}
Once \aaat finishes the pretraining process, it can be used as a speech editing system directly with an external duration predictor, which is shown in Fig.~\ref{fig:speech_edit}.

Given a speech $\vecs$, its original phonemes 
$\tilde{\vecx}$,
and the target modified phonemes $\vecx$,
our system first finds the phonemes that need to be modified $\hat{\vecx}$.
To predict the duration $\hat{\vecd_i}$ of modified phonemes
$\vecx_i \in \hat{\vecx}$, we use an external duration predictor.
Since the duration predictor only takes phoneme sequence as input,
we adjust model predicted duration $\vecd'_i$ according to the origin speech durations as:
$    \hat{\vecd}_i = \vecd_i^{'} \sum_{j=1}^{|\vecx|} {\tilde{\vecd}_j \over \vecd_j^{'}}$,
where $\tilde{\vecd_j'} / \vecd_j'$ is the ratio between
the duration of the original phone $\tilde{\vecx}_j$ in the given speech
and the predicted duration of the original phone $\tilde{\vecx}_j$
in the original sentence $\tilde{\vecx}$.
We compute this ratio from all phones in the original sentence $\tilde{\vecx}$
and use it
to adjust the predicted duration $\vecd'_i$ and get the final
duration $\hat{\vecd}_i$ for phone $\hat{\vecx}_i$.

With this predicted duration, we insert
$\sum_{i=1}^{|\tilde{\vecx}|} \hat{\vecd}_i \cdot \mathtt{sr} / \mathtt{h}$\footnote{$\mathtt{sr}$ stands for sample rate and $\mathtt{h}$ stands for hop size.}
number of \texttt{[MASK]} frames into the non-modified spectrogram context.
This masked spectrogram $\hat{\vecs}$ is the input of \aaat and 
$f ([e_{\hat{\vecs}}; \vecx]) + g \big( f ([e_{\hat{\vecs}}; \vecx])\big)$
is the prediction of the modified portion spectrogram.
Then, we use a vocoder to generate the waveform of this spectrogram
and output the final edited speech by replacing the modified part of the original speech.

\begin{figure}[t!]
\centering
\begin{tabular}{c}

\begin{minipage}[b]{0.9\linewidth}
    \begin{center}
    \subfigure[Speaker embedding-based method.]{
        \includegraphics[width=0.65 \linewidth]{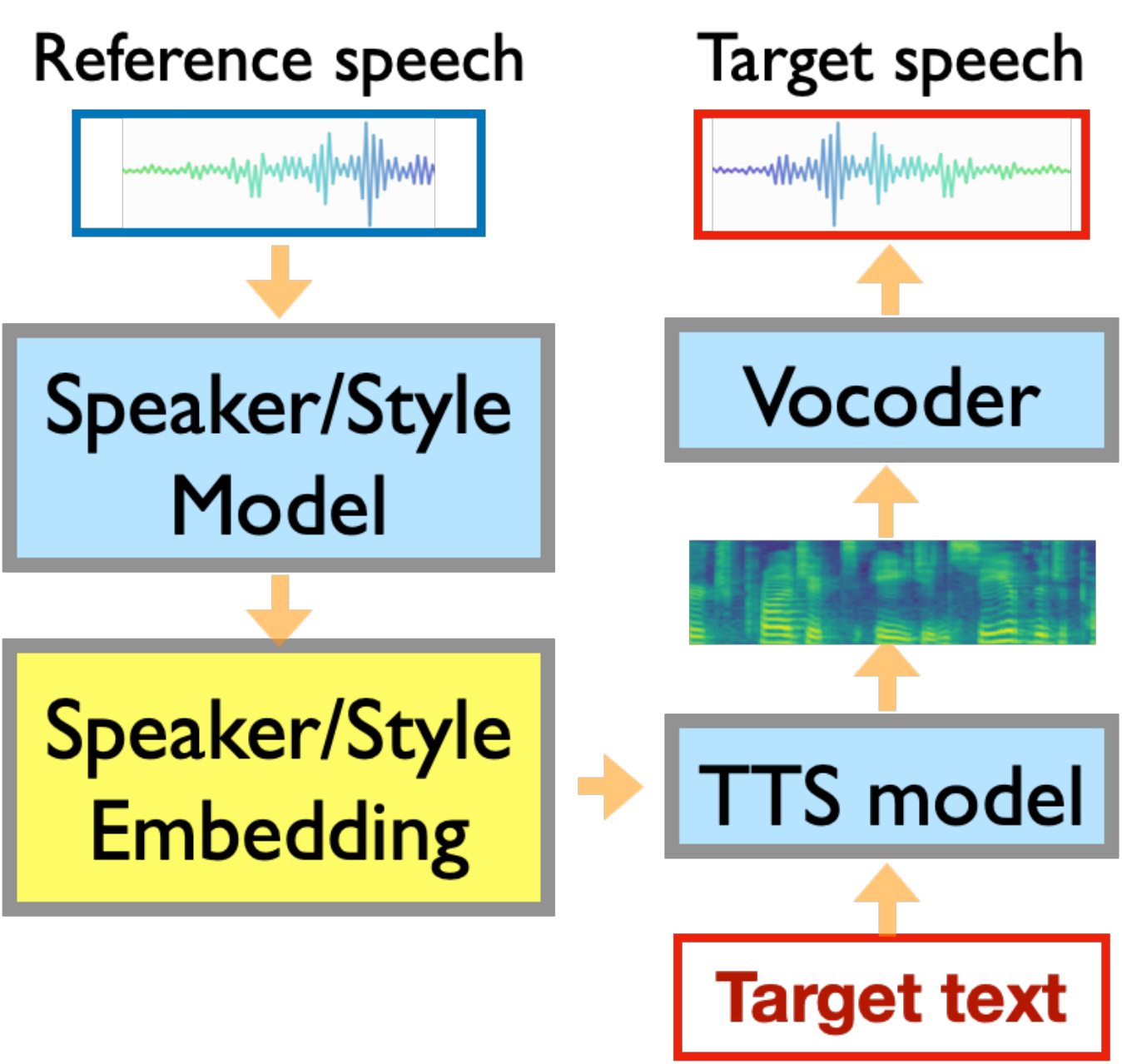}
        \label{fig: one-shot-tts-baseline}
    }
    \end{center}
    \end{minipage}
    \\

\begin{minipage}[b]{0.85 \linewidth}
    \begin{center}
    \subfigure[Prompt-based decoding.]{
        \includegraphics[width=1.0 \linewidth]{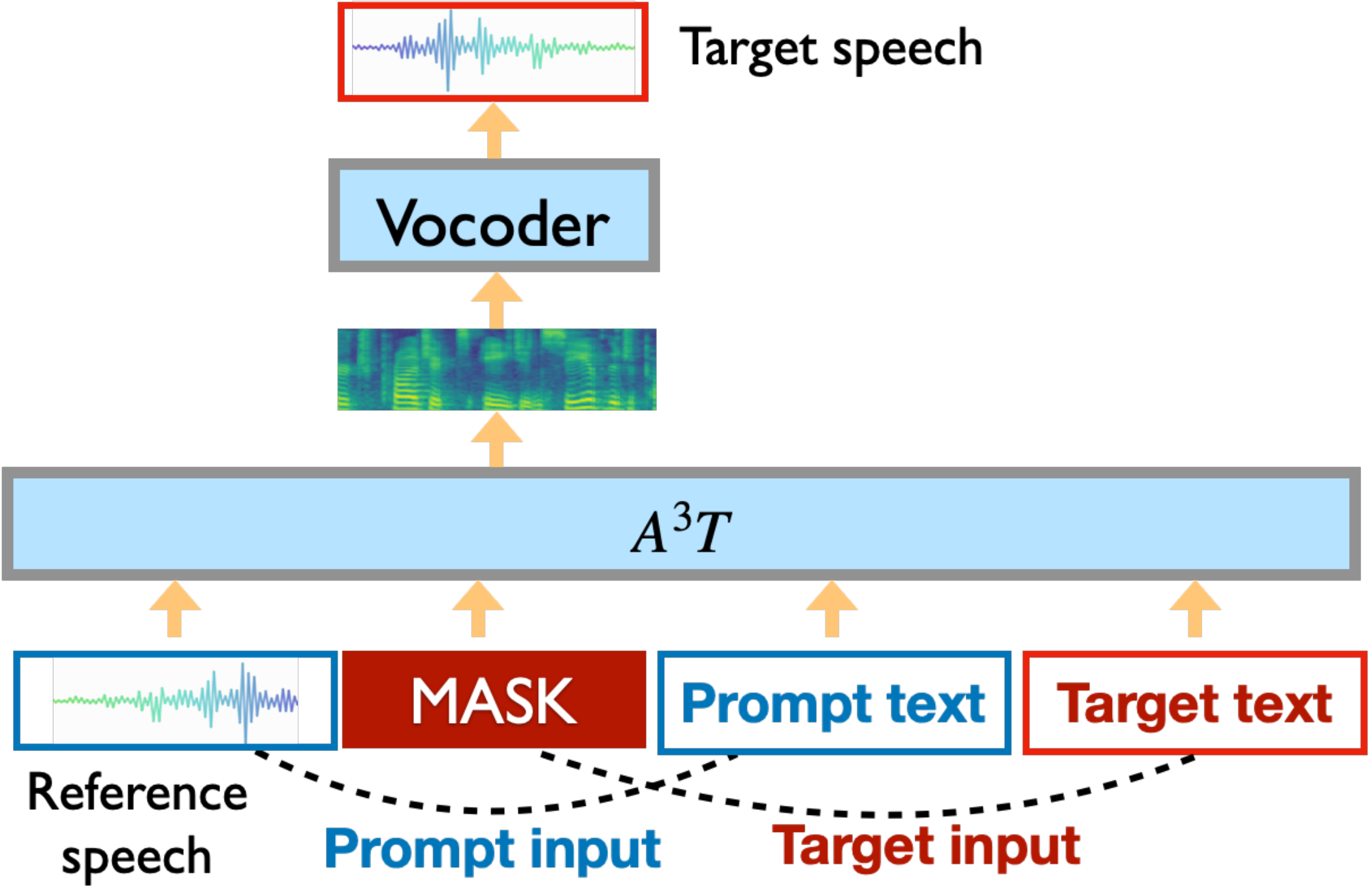}
        \label{fig: one-shot-tts-a3t}
    }
    \end{center}
    \end{minipage}
\end{tabular}
\caption{
Illustrations for one-shot TTS. The prompt speech and text are wrapped with blue rectangles, and the target speech and text are wrapped with red.
}
\vspace{-0.5cm}
\label{fig: one-shot-tts}
\end{figure}

\subsection{\aaat for Multi-speaker TTS}

In addition to the speech editing, we find our model has the potential for unseen speaker TTS.

Existing popular unseen speaker TTS models~\cite{jia2018transfer} are trained with seen speaker embeddings and generalizes to unseen speaker embeddings during the inference. 
However, such speaker embeddings are extracted from an external speaker verification model which is trained with tens of thousands of speakers.

In this work, we find our model can achieve comparable naturalness to models with speaker embeddings for unseen speaker TTS task;
What's more, our generations are more similar to the unseen speaker's reference speech.
The illustrations of how to synthesis speech for unseen speakers with our \aaat model are shown in Fig.~\ref{fig: one-shot-tts}, which is named prompt-based \aaat.

The key idea is to concatenate the prompt and the target together into a new utterance input, where the target speech is consist of $n$ \texttt{[MASK]} and $n$ is predicted by a duration predictor.
By inputting the concatenated speech and text, \aaat model will predict the spectrogram of these masked frames.
The role of the reference text and speech in our model is similar to prompts in language model~\cite{brown2020language}, and hence we call it prompt-based decoding/generation.

\section{Experiments}



In this section, we introduce our experiments for spectrogram reconstruction pretraining task, speech-editing task, and multi-speaker TTS. 
The spectrogram reconstruction is our pretraining task, where we conduct ablation study to show the contributions of different components and also the effects of different masking rates.
The experiment settings of speech-editing are followed~\citet{tan2021editspeech}, where we deploy two speech-editing systems with two datasets and evaluate the Mel-cepstral distortion~(MCD) score and
human-annotated mean opinion score~(MOS)~\cite{chu2006objective} using Amazon Mechanical Turk.
The multi-speaker TTS experiments include seen speaker TTS and unseen speaker TTS evaluated with the MOS scores.

\subsection{Datasets}



Following~\citet{tan2021editspeech}, we conduct our speech-editing experiments with a single-speaker TTS dataset LJSpeech~\cite{ljspeech17} and a multi-speaker TTS dataset VCTK~\cite{yamagishi2019vctk}.
The LJSpeech dataset is a single-speaker dataset with 13K examples in 24 hours.
The VCTK dataset is a multi-speaker dataset with 109 speakers and 44K examples in 44 hours.
It should be noted that after finishing the pretraining process with LJSpeech or VCTK, our \aaat will be used as a speech-editing system without any further finetuning.

We test multi-speaker TTS task with VCTK dataset. 
For seen multi-speaker TTS, each speaker's examples would be split into train and test sets. 
For unseen multi-speaker TTS, the test set contains 10 speakers' examples, and the other 99 speaker's examples are used for training.

\subsection{Configuration Details}

Raw audio files are processed with 50 ms frame size and 12.5 ms frame hop with the Hann window function to extract 80-dimensional log-Mel filterbanks.
We use 24K sampling rate for VCTK and 22K for LJSpeech.
The forced alignment and G2P are both carried out by HTK~\cite{young2002htk} to convert English words to phones and align phones with audio segments.
For speech-editing systems and prompt-based TTS, we use the publicly available duration predictor from FastSpeech 2 implemented in ESPnet~\cite{inaguma2020espnet}.
We use Parallel-WaveGAN~\cite{yamamoto2020parallel} vocoder for all the systems.

All \aaat models pretrained in our experiments share the same architecture: 4 layers Conformer encoder, 4 layers Conformer decoder, and 5 layers Conv1d  Post-Net, with 2 heads multi-head attention in 384-dim. The convolution kernel sizes of the encoder and decoder are 7 and 31, respectively.
The shape of alignment embeddings is (500, 384), where we assume the number of phones will not exceed 500 for a single input.
The shape of input phone embeddings is (73, 384), and we use a ReLU~\cite{agarap2018deep} nonlinear layer to transform 80-dim log-Mel filterbanks features to 384-dim.
The total number of parameters is 67.7M.

\begin{figure*}[ht!]
\centering
\begin{tabular}{c}
\begin{minipage}[t]{.5 \linewidth}
\begin{center}
\subfigure[
Groundtruth spectrogram from LJSpeech. 
]{
    \makebox[0.9\linewidth][c]{
\includegraphics[height=4cm]{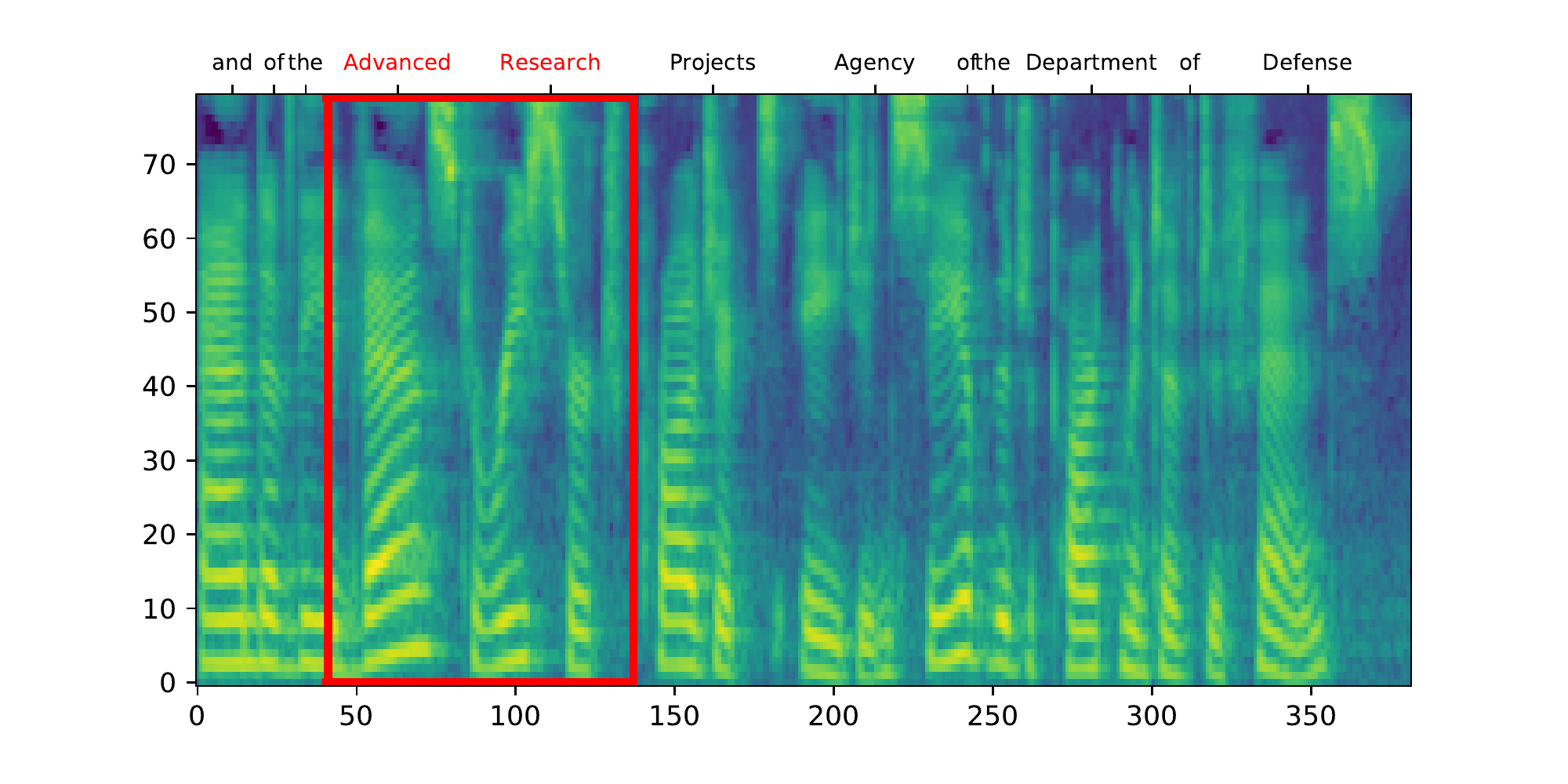}
\label{fig:ljs_ab_gt}
    }
}
\end{center}
\end{minipage}
\begin{minipage}[t]{.5 \linewidth}
    \begin{center}
    \subfigure[
    Reconstructed spectrogram by \aaat.
    ]{
    \makebox[0.9\linewidth][c]{
    \includegraphics[height=4cm]{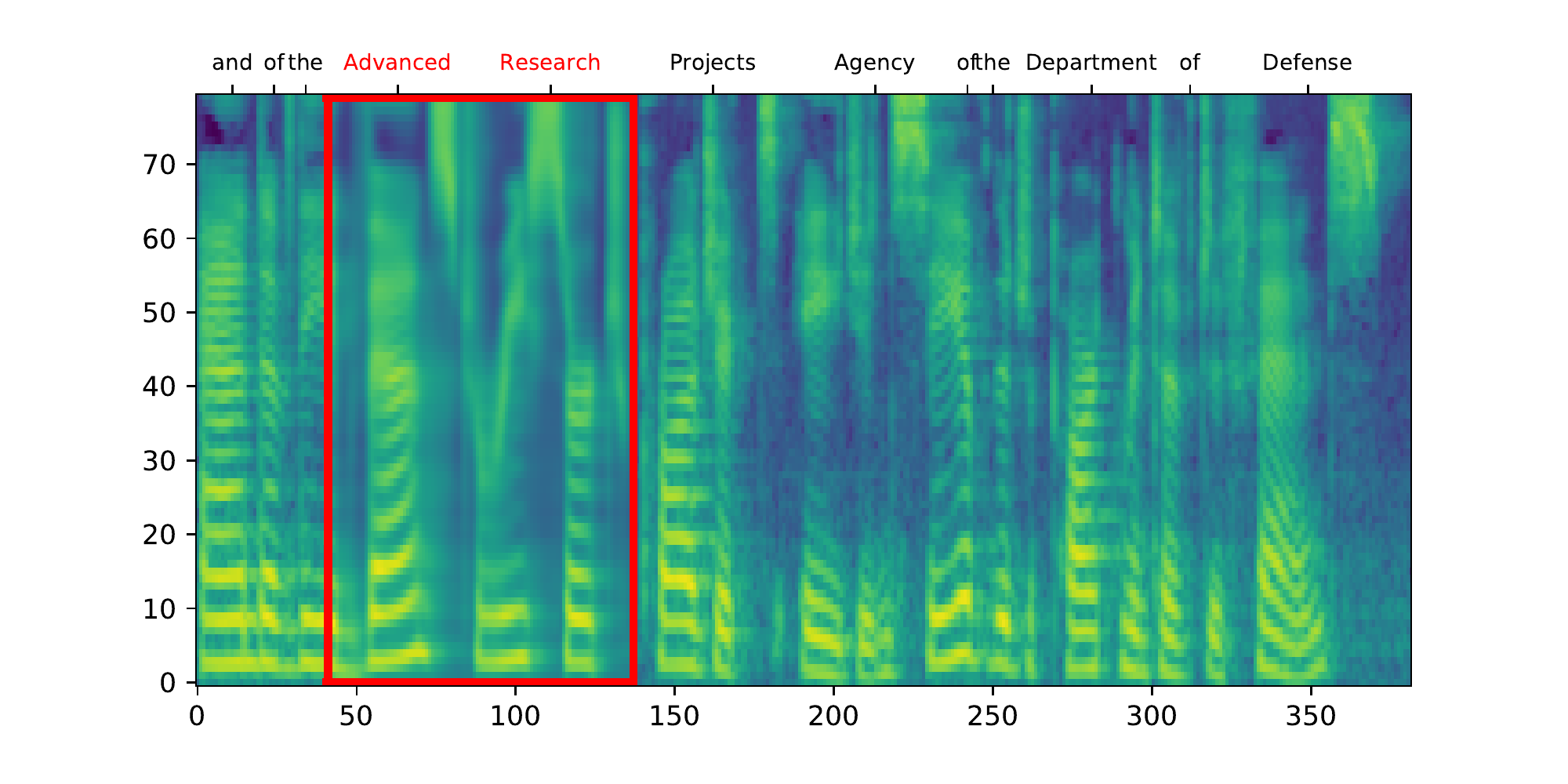}
    \label{fig:ljs_ab_b}
        }
    }
    \end{center}
\end{minipage}
    
\\

\begin{minipage}[t]{.5 \linewidth}
    \begin{center}
    \subfigure[
    Reconstructed spectrogram based on \aaat in (b) without segment embedding.
    ]{
    \makebox[0.9\linewidth][c]{
    \includegraphics[height=4cm]{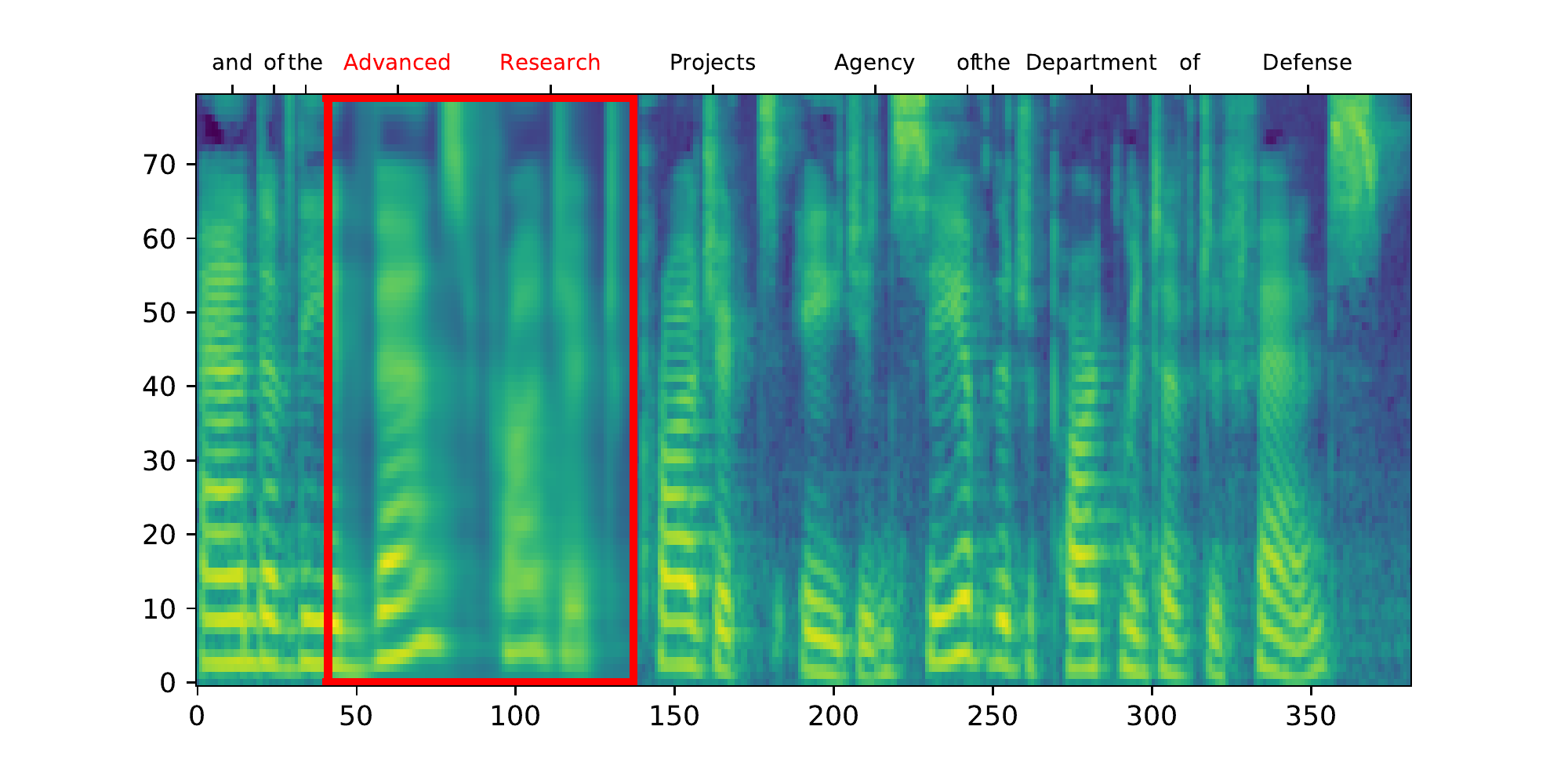}
    \label{fig:ljs_ab_c}
        }
    }
    \end{center}
\end{minipage}

\begin{minipage}[t]{.5 \linewidth}
    \begin{center}
    \subfigure[
    Reconstructed spectrogram based on \aaat in (c) with Transformer instead of Conformer.
    ]{
    \makebox[0.9\linewidth][c]{
    \includegraphics[height=4cm]{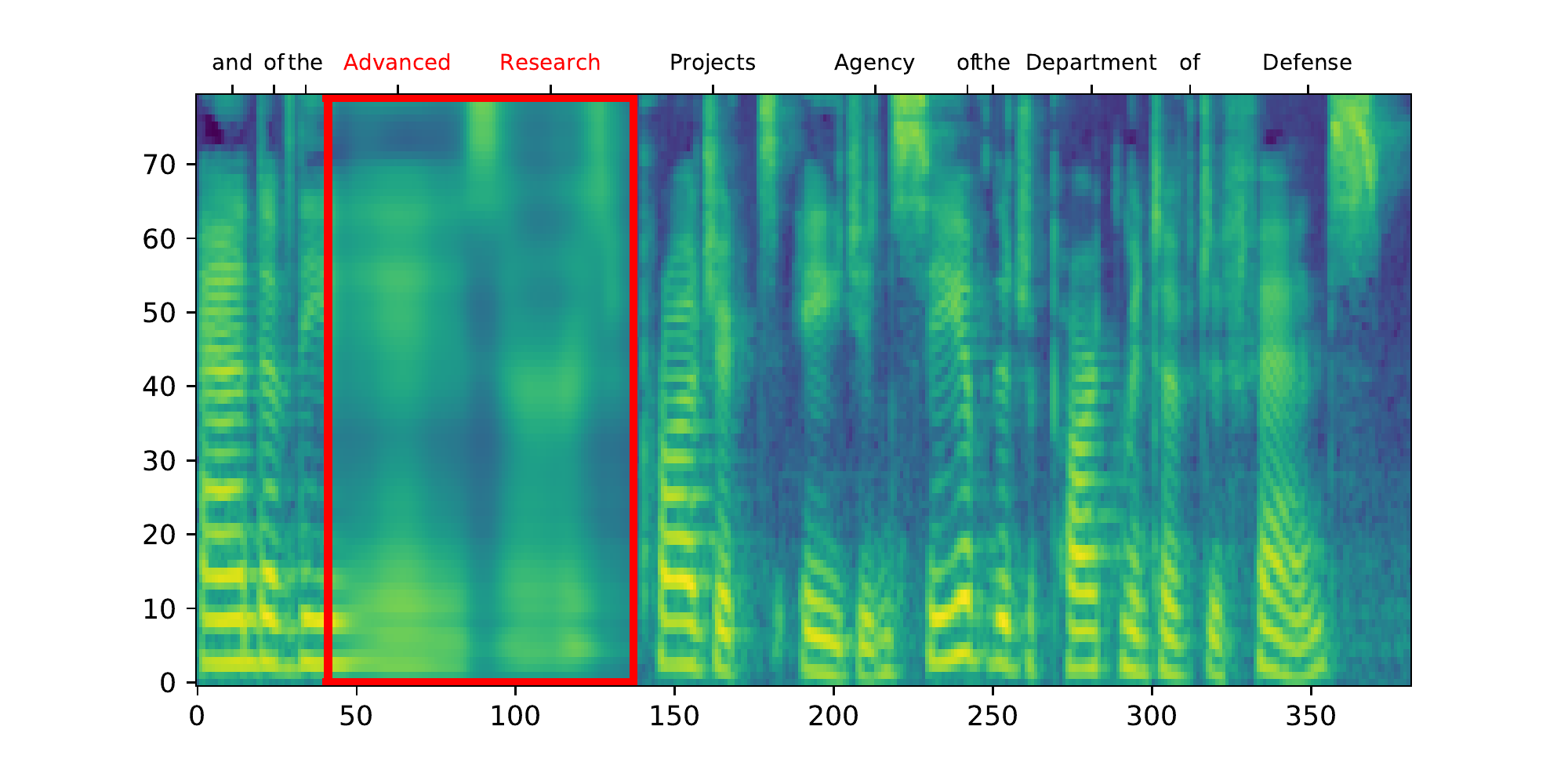}
    \label{fig:ljs_ab_d}
        }
    }
    \end{center}
\end{minipage}
\\

\begin{minipage}[t]{.5 \linewidth}
    \begin{center}
    \subfigure[
    Reconstructed spectrogram based on \aaat in (d) without Post-Net.
    ]{
    \makebox[0.9\linewidth][c]{
    \includegraphics[height=4cm]{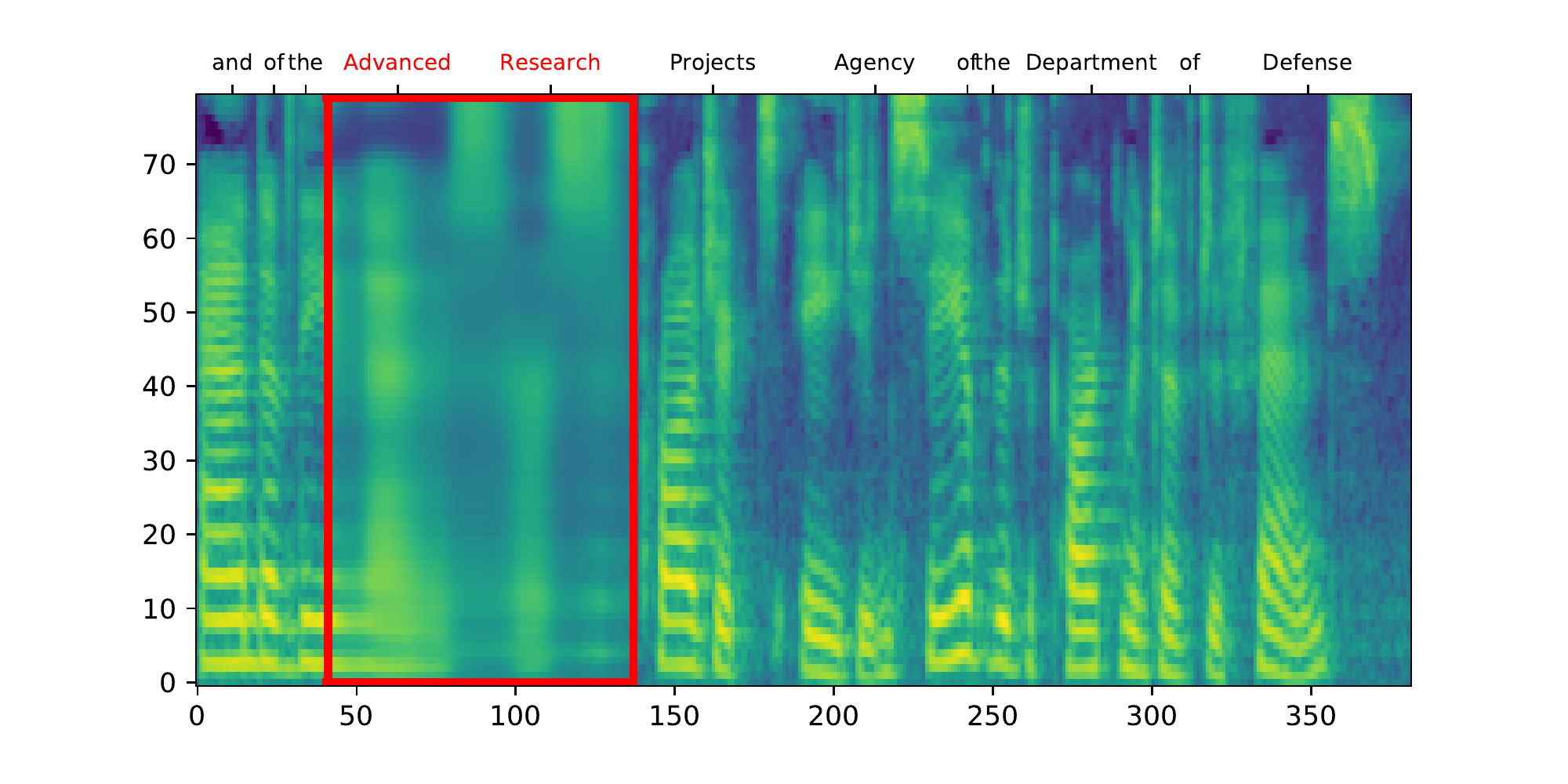}
    \label{fig:ljs_ab_e}
        }
    }
    \end{center}
\end{minipage}
    
\begin{minipage}[t]{.5 \linewidth}
    \begin{center}
    \subfigure[Reconstructed spectrogram based on \aaat in (e) with L2 loss instead of L1 loss. This model uses the similar achitecture of FAT-MLM.
    ]{
    \makebox[0.9\linewidth][c]{
    \includegraphics[height=4cm]{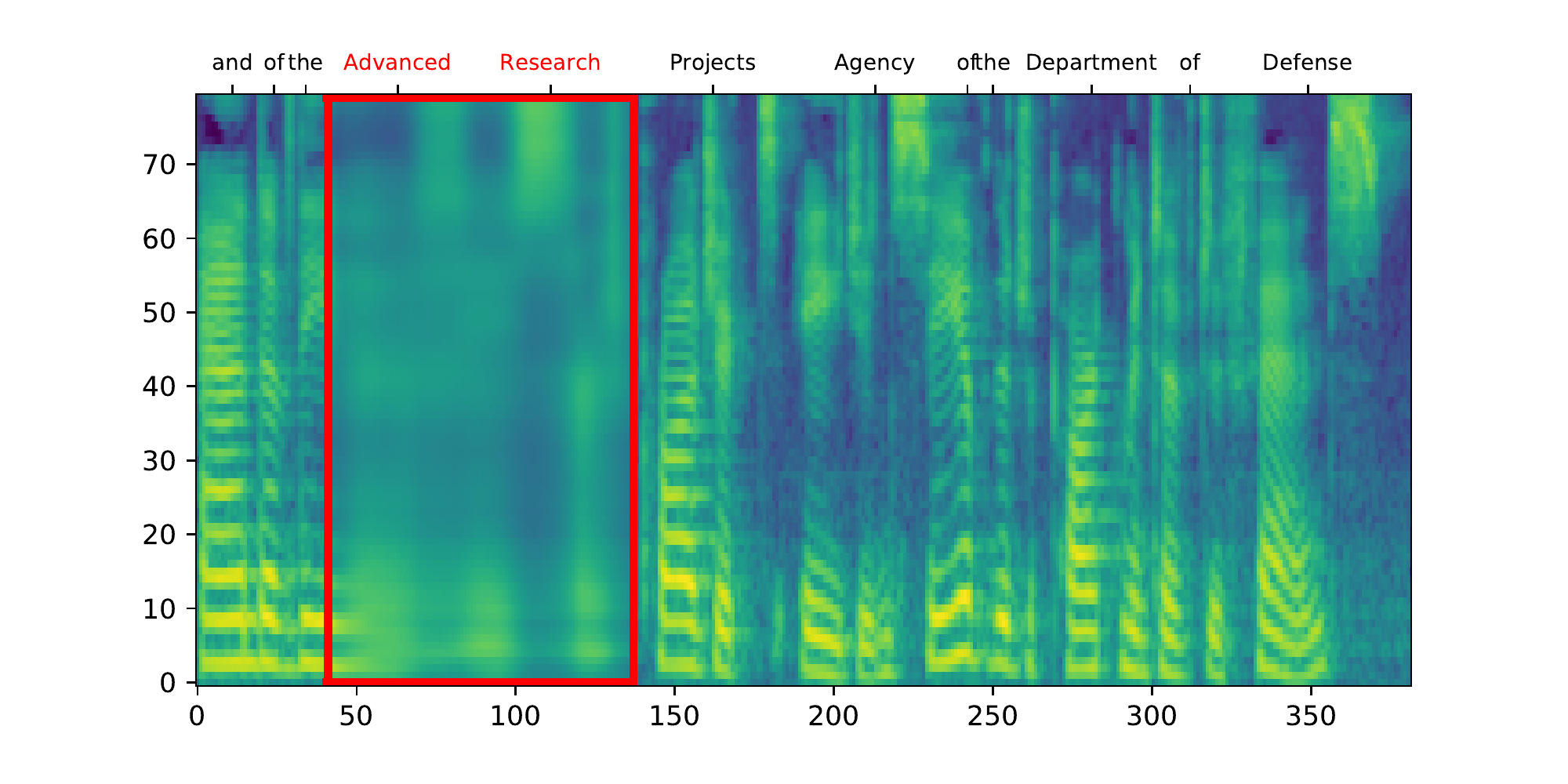}
    \label{fig:ljs_ab_f}
        }
    }
    \end{center}
\end{minipage}
    
\end{tabular}
\caption{
    An example of ablation study in LJSpeech.
    Original text is ``and of the \textcolor{red}{Advanced Research} Projects Agency 
    of the Department of Defense''. The portion with red box is 
    ``\textcolor{red}{Advanced Research}'' which is masked in (b,c,d,e,f) subfigures.
}
\label{fig:ablation}
\end{figure*}



During training, we use Adam optimizer with a 1.0 initial learning rate, 4000 warmup steps, and Noam learning rate scheduler. 
Instead of setting a fixed batch size, we adjust the batch size according to the length of the input example and set a maximum batch-bin~(the total number of input elements) for each model.
Following MAM~\cite{chen2020mam}, 15\% frames will be masked for speech-only input, 
For speech-text input, we randomly select several phonemes spans~( 80\% phonemes) and mask their corresponding frames.
For speech-editing experiments, we use 2.4M batch-bin, 1M steps for LJSpeech, and 3M batch-bin, 1.2M steps for VCTK. 

\begin{figure}[!t]
  \centering
\subfigure[Attention map of \aaat w/o alignment embeddings]{
\includegraphics[width=0.3\textwidth
]{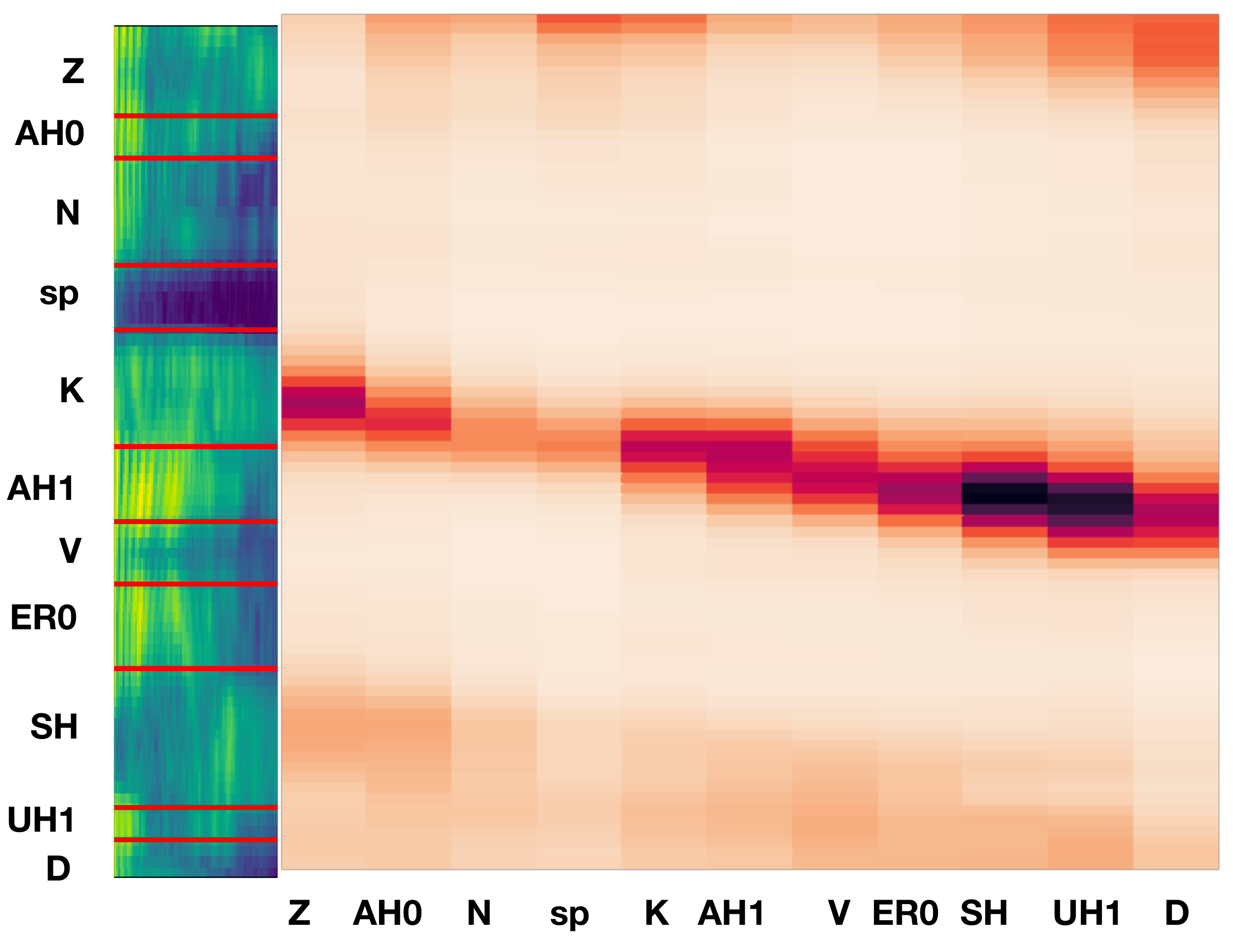}}
\\
\subfigure[Attention map of \aaat with alignment embeddings]{
\includegraphics[width=0.3\textwidth
]{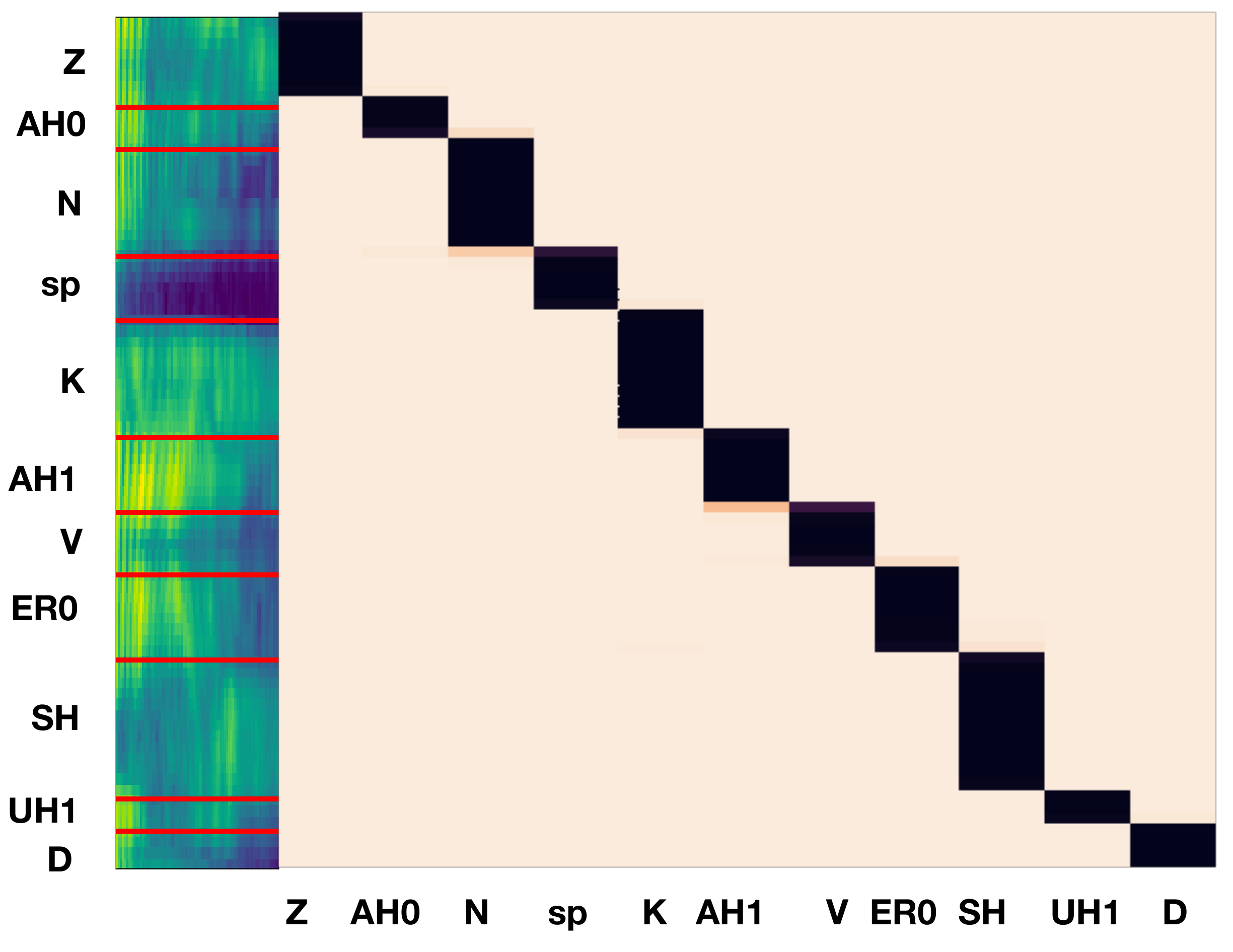}}
  \caption{Attention map between speech and text of \aaat with and without alignment embeddings.}
  \label{fig:attention_heatmap}
\end{figure}
    
\subsection{Ablation Study with Spectrogram Reconstruction}

\begin{table}[ht!]
	\centering
\resizebox{.9\columnwidth}{!}{
		\begin{tabular}{c | l |r}
\toprule
Example & \textbf{Model}  & \textbf{MCD} $\downarrow$ \\ 
	\midrule
Fig.~\ref{fig:ljs_ab_b} &  \aaat        &      8.09   \\ 
Fig.~\ref{fig:ljs_ab_c} &  \qquad  - Alignment Embeddings  &  10.73      \\
Fig.~\ref{fig:ljs_ab_d}  & \qquad  - Conformer  &     12.43    \\ 
Fig.~\ref{fig:ljs_ab_e}  & \qquad  - Post-Net  &     12.94   \\
Fig.~\ref{fig:ljs_ab_f}) & \qquad  - L1 loss  &     11.55    \\
\midrule
\end{tabular}
}
\caption{Ablation study for \aaat pretrained with LJSpeech. We remove modules one by one. For each model, we show a spectrogram example in Fig.~\ref{fig:ablation}. }\label{table:ablation}
\end{table}

\begin{table}[ht!]
	\centering
\resizebox{.8\columnwidth}{!}{
		\begin{tabular}{c|r|r}
\toprule

\textbf{MaskRate}  & \textbf{Seen MCD} $\downarrow$ & \textbf{Unseen MCD} $\downarrow$ \\ 
	\midrule
 20\%  &  10.35 & 12.22   \\ 
 50\%  &  7.75  & 9.99    \\
 80\%  &   8.72  & 9.68  \\
 	\midrule
\end{tabular}
}
\caption{MCD scores of \aaat pretrained in different masking rates with VCTK. }\label{table:ablation_masking_rate}
\end{table}

We first conduct an ablation study with LJSpeech dataset for our pretraining task: spectrogram reconstruction.
This task requires \aaat to predict the masked frames.
We sample 30 utterances randomly from the test set, and 1/3 phones in the middle of each sentence are masked.
We adopt MCD to measure the difference between the ground-truth audio and the reconstructed audio, where we only measure the MCD of the masked region an lower MCD means higher similarity.
We incrementally discard the components of \aaat: removing the cross-modal alignment embedding, replacing the Conformer with Transformer, removing the Post-Net, and using L2~(MSE) loss instead of L1~(MAE) loss.

Results are shown in Tab.~\ref{table:ablation}.
An example of different models' reconstruction is shown in Fig.~\ref{fig:ablation}.
By comparing Fig.~\ref{fig:ljs_ab_b} and Fig.~\ref{fig:ljs_ab_c}, we can see that many details are lost when \aaat trained without the alignment embedding, and the MCD scores rise from 8.09 to 10.73.
Similar degrading can be observed after replacing Conformer with Transformer: the MCD scores rise from 10.73 to 12.43 and the spectrogram becomes blurrier~(Fig.~\ref{fig:ljs_ab_d}).
Compared with the alignment embedding and Conformer, Post-Net contributes only 0.49 MCD score, and L2 loss even achieves better MCD score than L1 loss.
However, when looking into the spectrograms, we can see that Fig.~\ref{fig:ljs_ab_f} is blurrier than Fig.~\ref{fig:ljs_ab_e}, which conforms to the previous finding~\cite{klimkov2018parameter} that L1 loss is better than L2 loss for speech synthesis.
Hence, we choose L1 loss for \aaat pretraining.
Also, Fig.~\ref{fig:ljs_ab_f} indicates the quality that previous pretrained model~(MAM/FAT-MAM) could achieve, and the other figures show how our \aaat transforms Fig.~\ref{fig:ljs_ab_f} to Fig.~\ref{fig:ljs_ab_b}.

We also conduct a study with VCTK to show the impacts of difference masking rates.
Results are shown in Tab.~\ref{table:ablation_masking_rate}.
We can see that 20\% masking rate leads to large MCD scores, while 50\% and 80\% are better.
Also, 50\% masking rate outperforms 80\% on the seen test cases, but not on the unseen.
Considering 80\% masking rate has a better generalization on unseen cases, we choose 80\% for all the following experiments. 

Finally, we plot the attention heat maps of encoder with and without our proposed cross-modal alignment embedding in Fig.~\ref{fig:attention_heatmap}.
The attention matrices are collected from the encoder's last layer with a mean-pooling across heads.
It should be noted that the original attention matrix is 310*310, which contains both the speech and phones, and for clarity, we plot only 11 phones and their corresponding frames in Fig.~\ref{fig:attention_heatmap}.
We can see that our \aaat is aware of the speech segmentations and their corresponding phones, while the baseline model fails to capture such alignment information.
This observation demonstrates the effectiveness of our \aaat for cross-modal pretraining.
This observation also conforms previous finding that Transformer-based language model cannot align the tokens within the same sentnece/paragraph together, even pre-trained with the BERT-large setting~\cite{bai2021segatron}.


\begin{figure}[h!]
\centering
\begin{tabular}{c}
\begin{minipage}[t]{.3 \linewidth}
\begin{center}
\subfigure[
Baseline 1.
]{
    \makebox[.85\linewidth][c]{
\includegraphics[height=3.5cm]{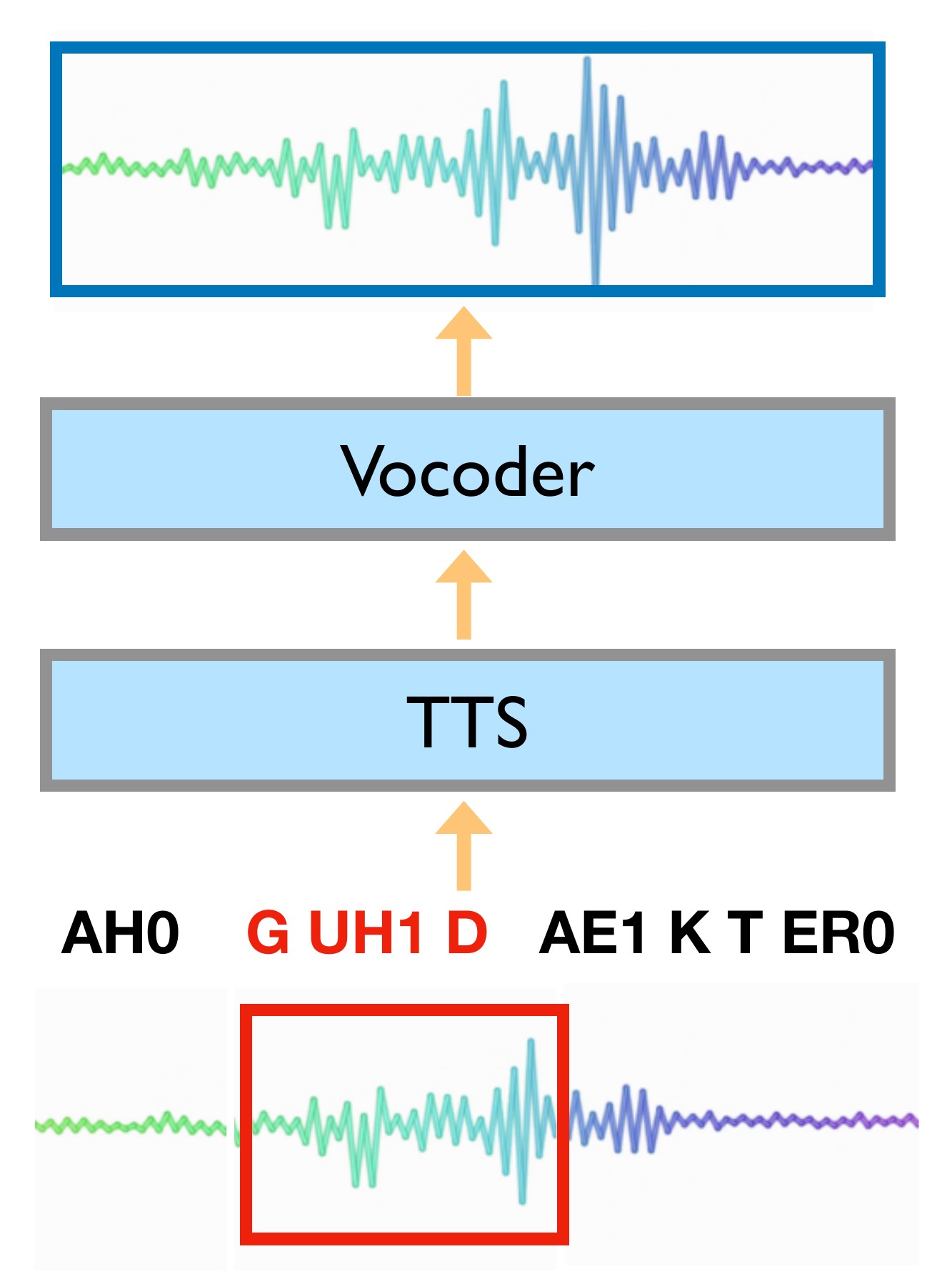}
\hspace{10pt}
\label{fig: edit_baseline1}
    }
    
    \begin{tikzpicture}
    \draw[gray, dashed] (0,0) -- (0,3.5);
    \end{tikzpicture}
}
\end{center}
\end{minipage}
\begin{minipage}[t]{.35 \linewidth}
\begin{center}
\subfigure[
Baseline 2.
]{
\hspace{-8pt}
\makebox[.85\linewidth][c]{
\includegraphics[height=3.5cm]{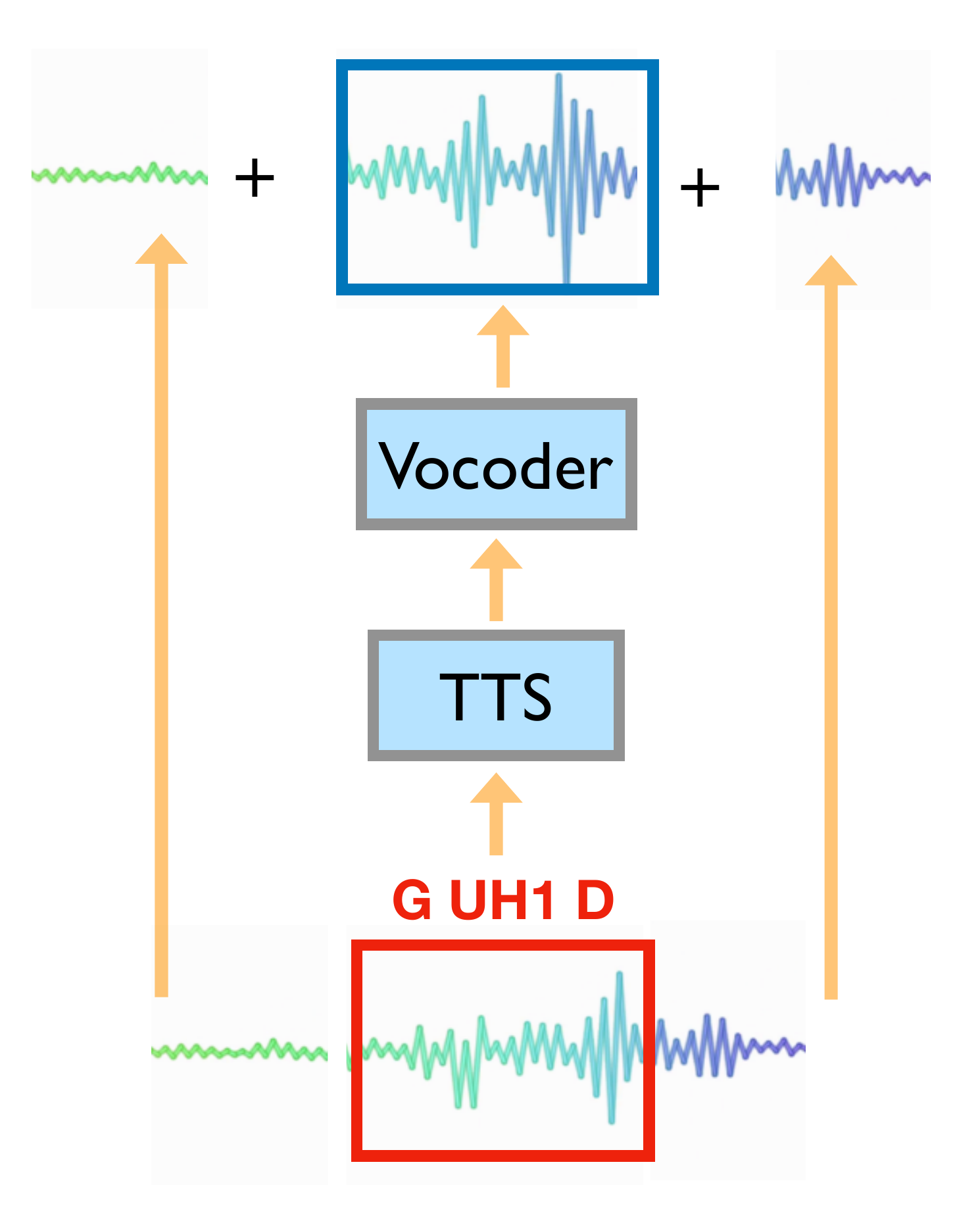}
\hspace{4pt}
\label{fig: edit_baseline2}
    }
    \begin{tikzpicture}
    \draw[gray, dashed] (0,0) -- (0,3.5);
    \end{tikzpicture}
}
\end{center}
\end{minipage}
\begin{minipage}[t]{.3 \linewidth}
\begin{center}
\subfigure[
Baseline 3.
]{

\makebox[.85\linewidth][c]{
\includegraphics[height=3.5cm]{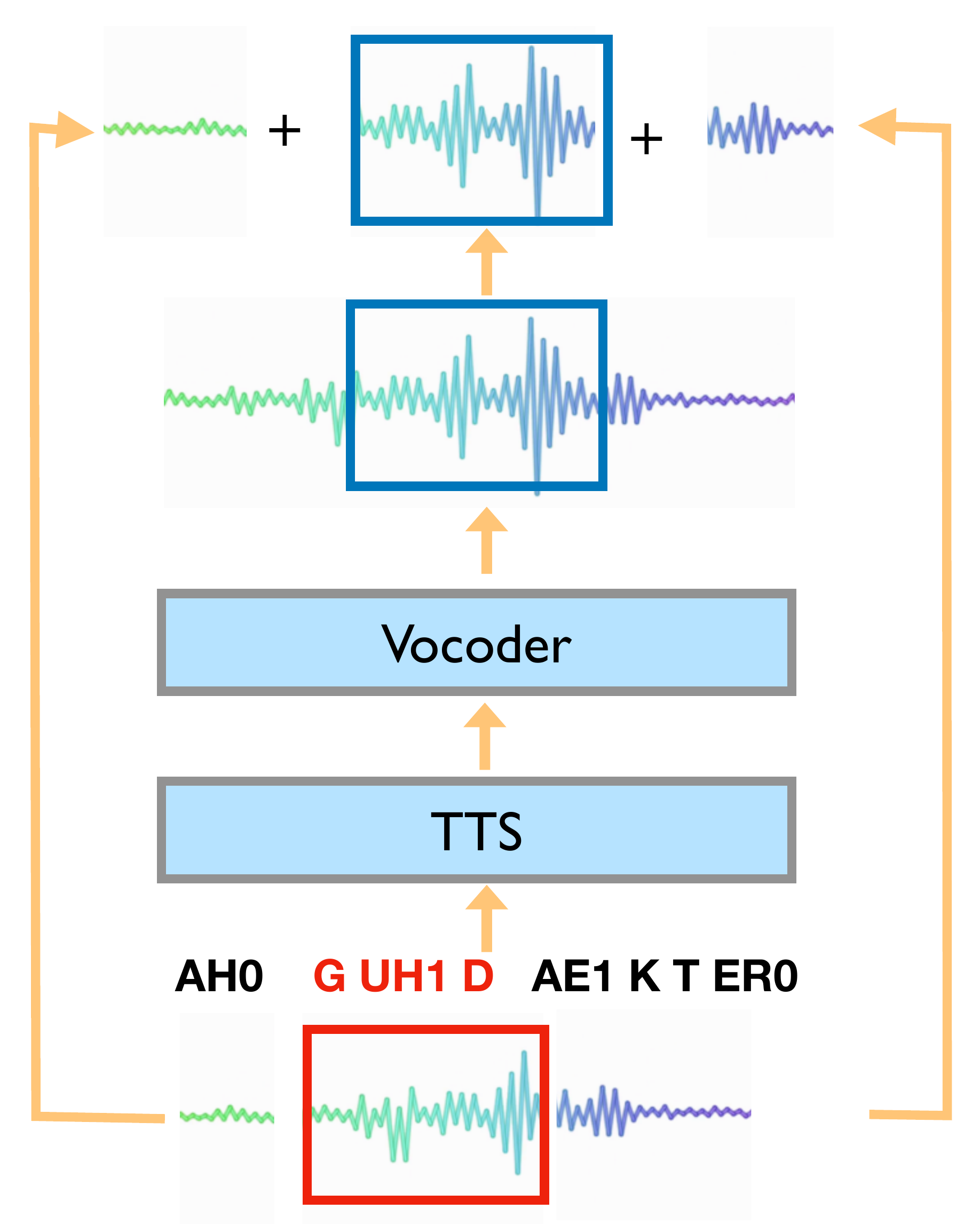}
\label{fig: edit_baseline3}
    }
}
\end{center}
\end{minipage}
\end{tabular}
\caption{
Illustrations for speech editing baselines.
}
\vspace{-0.3cm}
\label{fig:baselines}
\end{figure}
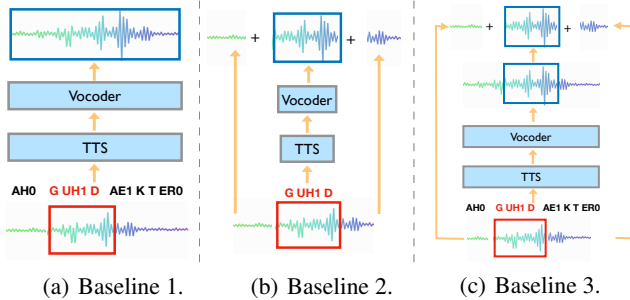

\subsection{Speech Editing}
\begin{figure}[t!]
    \centering
    \begin{tabular}{c}
    \begin{minipage}[b]{1\linewidth}
        \begin{center}
        \subfigure[\aaat train/decode]{
            \includegraphics[width=0.4 \linewidth]{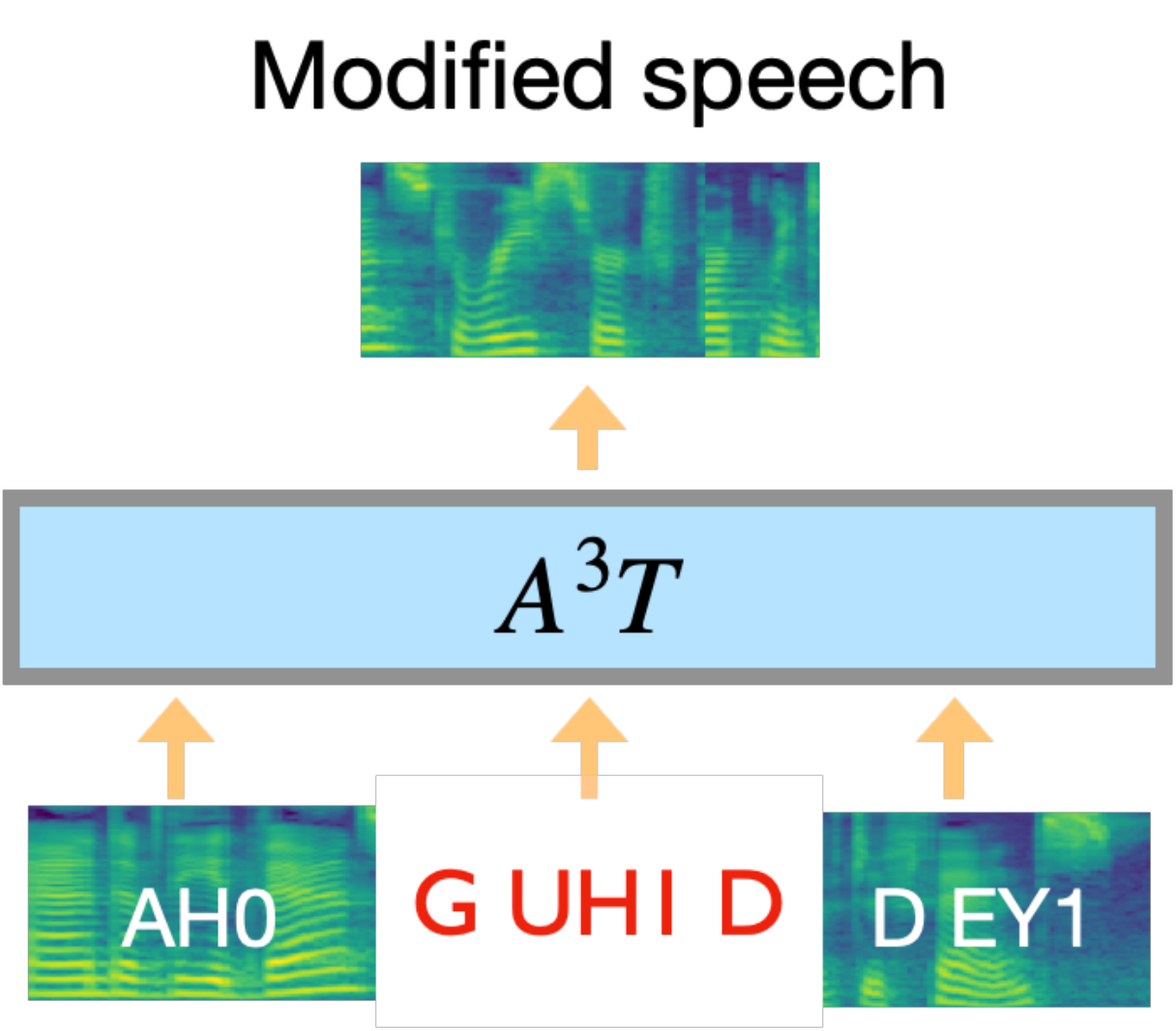}
            \label{fig: a3t_both}
        }\hfill
        \subfigure[EditSpeech train]{
            \includegraphics[width=0.35 \linewidth]{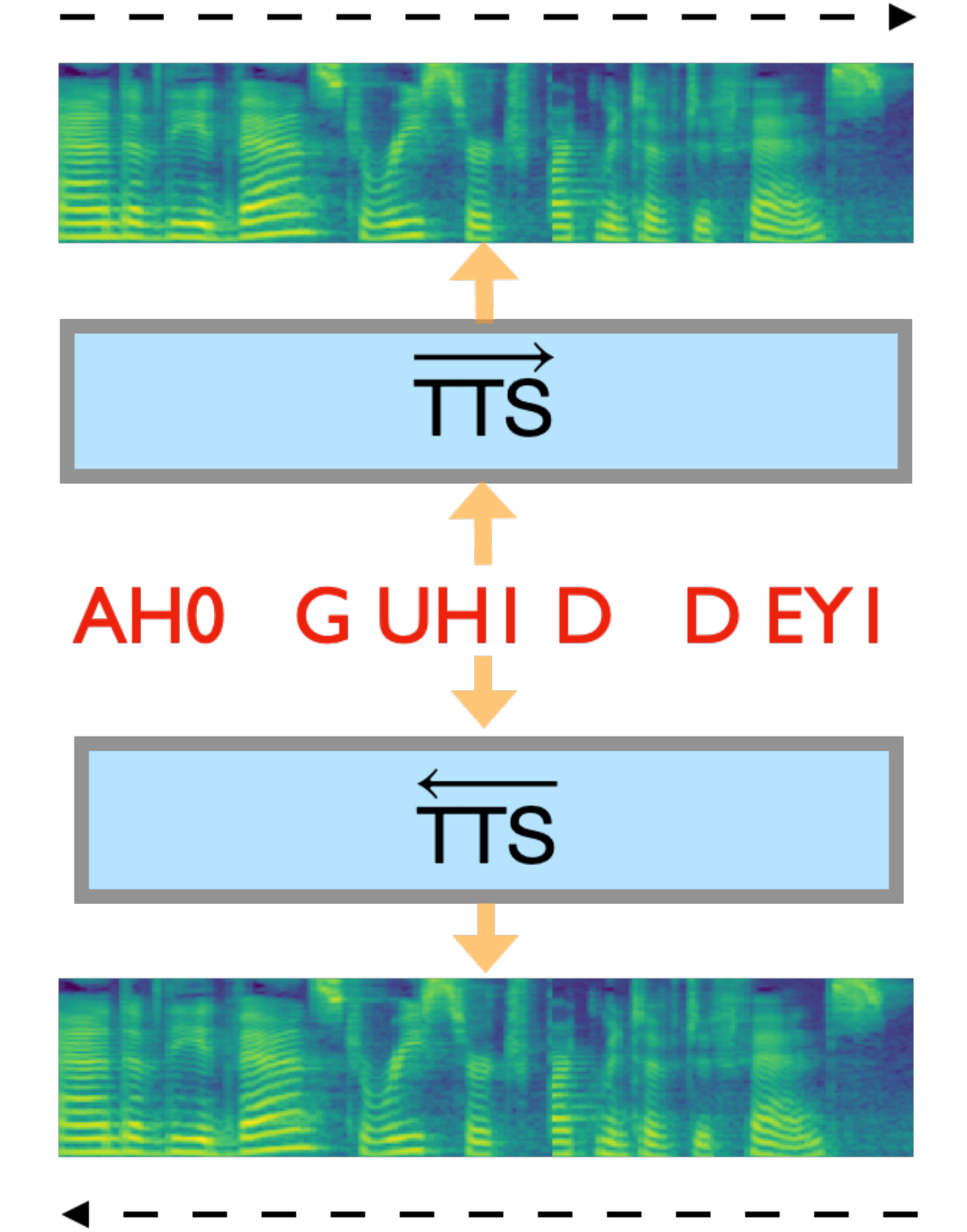}
            \label{fig: es_train}
        }\\
        \subfigure[EditSpeech decode]{
            \includegraphics[width=0.75 \linewidth]{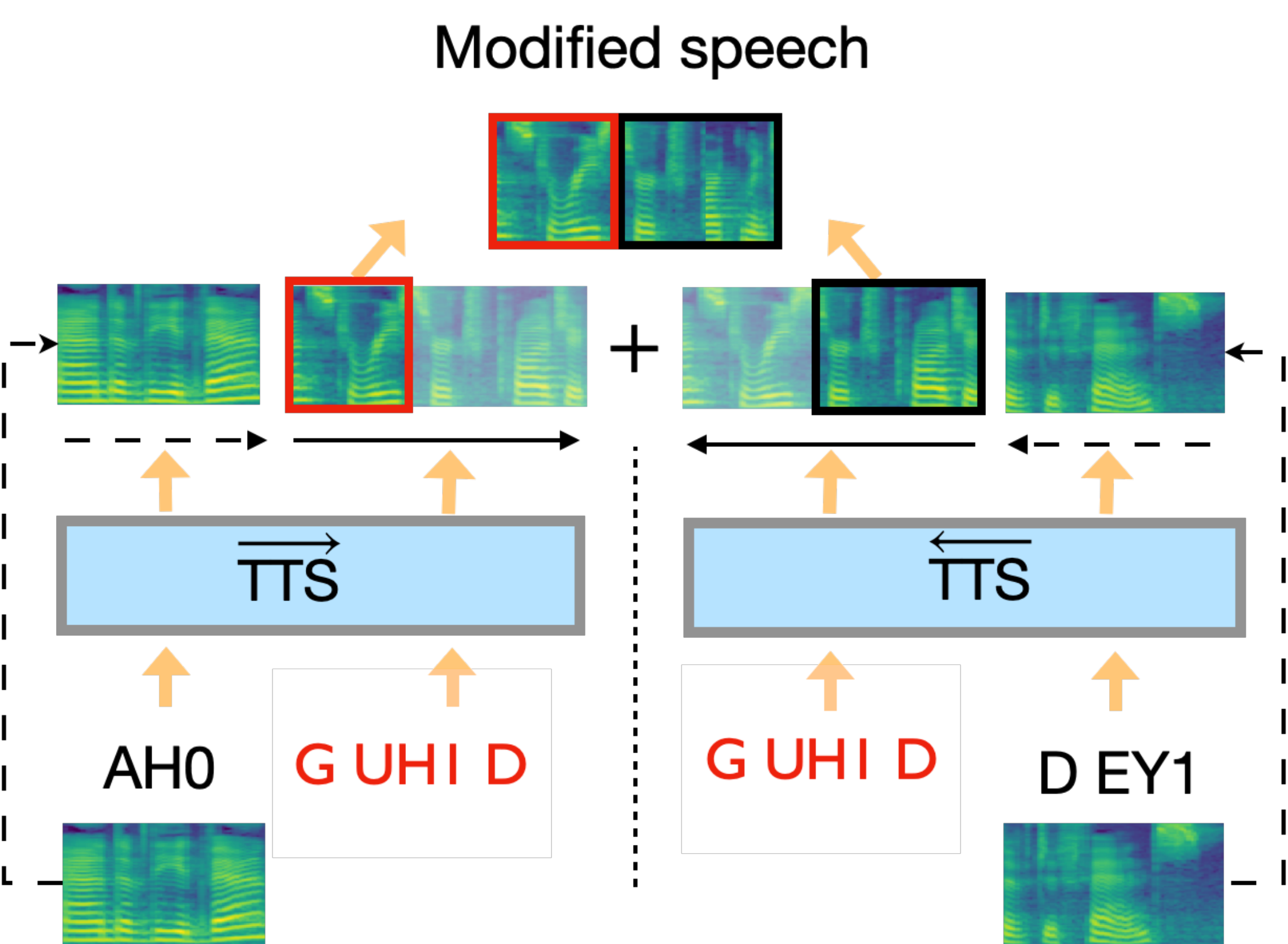}
            \label{fig: es_decode}
        }
        \end{center}
        \end{minipage}
    \end{tabular}
    \caption{
    Comparisons between \aaat and EditSpeech. $\rightarrow$: free decoding, $\dashrightarrow$: forced decoding. \aaat trains a non-autoregressive encoder
    to reconstruct masked acoustic signals, and uses the identical framework for decoding. 
    EditSpeech trains two autoregressive TTS models: a left-to-right and a right-to-left.
    For decoding, these two models synthesize two speeches and are fused for the output.
    }
    \label{fig:baselines}
    \end{figure}

Following \citet{tan2021editspeech}, we list several baseline systems below:

\begin{itemize}

\item \textbf{Baseline 1}: This is a TTS system regenerating
a complete waveform from the whole sentence to be edited.
\item \textbf{Baseline 2}: This system generates the modified region with a TTS model and insert the generation back to the original waveform with a forced aligner.
\item \textbf{Baseline 3}: This system is similar to Baseline 1, but we cut the modified region from the generation and insert it back to the original waveform with a forced aligner. 
\item \textbf{\citet{tan2021editspeech}}: This is a speech-editing system which introduces partial inference and bidirectional fusion to sequence-to-sequence neural TTS model. 
EditSpeech trains two conventional autoregressive TTS models,
one left-to-right and the other right-to-left 
(Fig.~\ref{fig: es_train}).
For decoding, the left-to-right TTS model 
force-decodes the prefix speech context and synthesizes the modified region, and the right-to-left TTS model force-decodes the suffix context and generates the modified region reversely.
Finally, the two synthesized speeches are fused for final output (Fig.~\ref{fig: es_decode}).

\end{itemize}

Following \citet{tan2021editspeech}, we also evaluate our speech editing system with an identical reconstruction task, which is similar to the above ablation experiments but without the ground-truth duration length and can be evaluated with MCD metric.
30 utterances are randomly sampled for each dataset, and a part of speech, which corresponds to 1/3 phonemes in the middle of each sentence, is masked.
The audio of the masked region is replaced with each system's generation.
A duration model is used to predict the length of
masked speech from phonemes.
Results are shown in Tab.~\ref{table:mcd}. 
From this table, we can see that our system achieves the best MCD score.
Besides, alignment embedding is the key to reducing MCD, which confirms our observation in Fig.~\ref{fig:ljs_ab_c}.
For TTS-based systems, we find that generating the whole audio and then extracting the modified region is better than generating the modified region only.

We then conduct the human evaluation with Amazon Mechanical Turk
for the real speech insertion and replacement tasks using the VCTK dataset.
To compare our results with~\citet{tan2021editspeech}, 
we use the same 15 audio samples and modification operations from their work.
For each audio sample, we use 10 English native speakers to evaluate the naturalness
of synthesized audios.
In Tab.\ref{table:sedit}, our \aaat speech editing system outperforms~\citet{tan2021editspeech}'s and gets the highest MOS scores among all these systems.
Audio examples can be found at our demo link.
\begin{table}[ht!]
	\centering
	\begin{subtable}{
\resizebox{0.9\columnwidth}{!}{
		\begin{tabular}{l | r | r}
\toprule
Model    & VCTK MCD $\downarrow$    & LJSpeech MCD $\downarrow$     \\ 
\midrule
Baseline 1/3 & 10.66      & 10.32           \\ 
Baseline 2 & 12.06    & 10.91             \\ 
\aaat      & \textbf{7.76}      & \textbf{9.26}            \\ 
\qquad w/o Alignment Emb. & 11.37      & 10.30     \\ 
\midrule
\end{tabular}}
}
    \end{subtable}
	\caption{MCD evaluation on identity speech reconstruction using VCTK and LJSpeech.}\label{table:mcd}
\end{table}

\begin{table}[ht!]
	\centering
\resizebox{0.8\columnwidth}{!}{
		\begin{tabular}{l|c|c}
\toprule
Model  & Insert & Replace \\ 
	\midrule
Baseline 1   & 3.02 $\pm$ 0.20   &   2.64 $\pm$ 0.16  \\ 
Baseline 2   & 2.89 $\pm$ 0.17   &  2.70 $\pm$ 0.16 \\
Baseline 3   & 2.89 $\pm$ 0.17   &  2.44 $\pm$ 0.16   \\
\citet{tan2021editspeech} & 3.50 $\pm$ 0.16 & 3.58 $\pm$ 0.16     \\
\aaat      & \textbf{3.53} $\pm$ 0.17 &  \textbf{3.65} $\pm$ 0.15    \\
\qquad w/o Alignment Emb.  & 2.48 $\pm$ 0.21   &  1.98 $\pm$ 0.17   \\
\midrule
\end{tabular}

}
\caption{The MOS evaluation ($\uparrow$) on speech editing task on VCTK with 95\% confidence intervals.}\label{table:sedit}
\end{table}

\begin{table}[ht!]
	\centering
\resizebox{0.8\columnwidth}{!}{
		\begin{tabular}{l|c|c}
\toprule
Model  & Seen & Unseen \\ 
	\midrule
FastSpeech 2 & 3.33 $\pm$ 0.10 & 3.78 $\pm$ 0.10 \\
+GST~\cite{wang2018style} & 3.42 $\pm$ 0.10 & 3.81 $\pm$ 0.11 \\
\aaat        & 3.61 $\pm$ 0.09 & 3.90 $\pm$ 0.10 \\
Groundtruth  & 3.94 $\pm$ 0.08 & 4.09 $\pm$ 0.10 \\
\midrule
\end{tabular}

}
\caption{The MOS evaluation ($\uparrow$) for \textbf{speaker similarity} on multi-speaker TTS on VCTK with 95\% confidence intervals. The FastSpeech2 model is equipped with X-vectors~\cite{snyder2018x}. }\label{table: vctk_similarity}
\end{table}

\begin{table}[ht!]
	\centering
\resizebox{0.8\columnwidth}{!}{
		\begin{tabular}{l|c|c}
\toprule
Model  & Seen & Unseen \\ 
	\midrule
FastSpeech 2 & 3.34 $\pm$ 0.11 & 3.85 $\pm$ 0.11 \\
+GST~\cite{wang2018style} & 3.27 $\pm$ 0.11 & 3.72 $\pm$ 0.11 \\
\aaat        & 3.63 $\pm$ 0.10 & 3.94 $\pm$ 0.11 \\
Groundtruth  & 4.04 $\pm$ 0.08 & 4.05 $\pm$ 0.10 \\
\midrule
\end{tabular}

}
\caption{ The MOS evaluation ($\uparrow$) for \textbf{speech quality} on multi-speaker TTS on VCTK with 95\% confidence intervals. The FastSpeech2 model is equipped with X-vectors~\cite{snyder2018x}. }\label{table: vctk_quality}
\end{table}

\subsection{Prompt-based Multi-speaker TTS}
We also conduct the human evaluation for multi-speaker TTS systems with seen speaker 
(30 test cases, 15 human annotations for each test case) 
and unseen speaker (20 test cases, 15 human annotations for each test case) testing cases.
The quality of the generations and the speaker similarity between the generation and the reference are evaluated,
and the results are shown in Tab.~\ref{table: vctk_similarity} and Tab.~\ref{table: vctk_quality}.
From this table, we can see that the style embedding GST~\cite{wang2018style} improves 
the similarity scores but harms the quality scores, 
while our \aaat model is the most favorable system in both the speaker similarity and the speech quality.
Strikingly, we observe that the average score of the Unseen cases is higher than the Seen, 
which is counterintuitive.
However, when looking into the MOS of the Groundtruth, 
the gap is still there and we believe this is due to the difference between these two test case sets.
\section{Discussion}
\textbf{\aaat is a pretraining method on parallel data.} 
\aaat is 
a BERT-style pretraining method,
which takes both phonemes and partially-masked spectrograms 
as inputs (Fig.~\ref{fig: a3t_both}). 
Although \aaat can be first trained with speech-only data~(Appendix~\ref{appendix: tts finetuning}), but a second stage of training with parallel data is necessary for speech synthesis.
\aaat trains a non-autoregressive encoder
to reconstruct masked acoustic signals, 
and uses the identical framework for decoding.
It is therefore akin to cross-lingual BERT like XLM~\cite{lample2019cross}
which also trains on parallel data.

\textbf{Finetuning.} 
In this paper, our major finding is that \aaat can be directly used without finetuning~(Tab~\ref{table:ablation}-~\ref{table: vctk_quality}), like GPT-3, for downstream tasks such as speech editing~(Tab.~\ref{table:sedit}) and unseen-speaker TTS~(Tab.~\ref{table: vctk_quality}).
We also find \aaat can be pretrained with more data and be finetuned, like BERT, to improve downstream tasks. 
Finetuning results for multispeaker TTS are reported in Appendix~\ref{appendix: tts finetuning}. 

\textbf{\aaat is not a TTS model.}
 The input of \aaat mush be both the text and the speech context, while the traditional TTS models' input is only the text. 
 We show a synthesized speech example in our demo, whose input is the text and a piece of silent audio. We find the generated speech sounds like multiple speakers are speaking the text simultaneously. This observation shows that \aaat generates speech based on the given context and follows its properties.
 On the other hand, \aaat can become a TTS model after finetuned with TTS task and data, which is introduced in the Appendix~\ref{appendix: tts finetuning}.

\section{Conclusions}

In this paper, we propose 
Alignment-Aware Acoustic-Text Pretraining 
(A$^3$T) which can reconstruct
masked acoustic signals with high quality.
We show that our proposed A$^3$T model has the
ability to do speech editing and outperforms the current
SOTA models,
and also improves unseen-speaker speech synthesis with our proposed prompt-based decoding.






\balance
\bibliography{main}
\bibliographystyle{icml2022}


%



\end{CJK}
\clearpage
\newpage
{\LARGE\textbf{Appendix}}
\appendix
In this appendix, we first show the generated spectrogram comparisons between our system and \citet{tan2021editspeech}'s system in Sec.~\ref{appendix: spectro}.

Then, we show our \aaat can also be used for TTS finetuning in Sec.~\ref{appendix: tts finetuning}. As experiments in Sec.~\ref{appendix: tts finetuning} are using more training steps or more data, we decide to put these experiments in the appendix instead of the main text for readers interested in the finetuning.

\section{Spectrogram Comparison of Speech Editing}
\label{appendix: spectro}
\begin{figure}[hb!]
\centering
\begin{tabular}{c}
\begin{minipage}[t]{.7 \linewidth}
\begin{center}
\subfigure[
Groundtruth spectrogram from VCTK.
]{
    \makebox[\linewidth][c]{
\includegraphics[height=4.5cm]{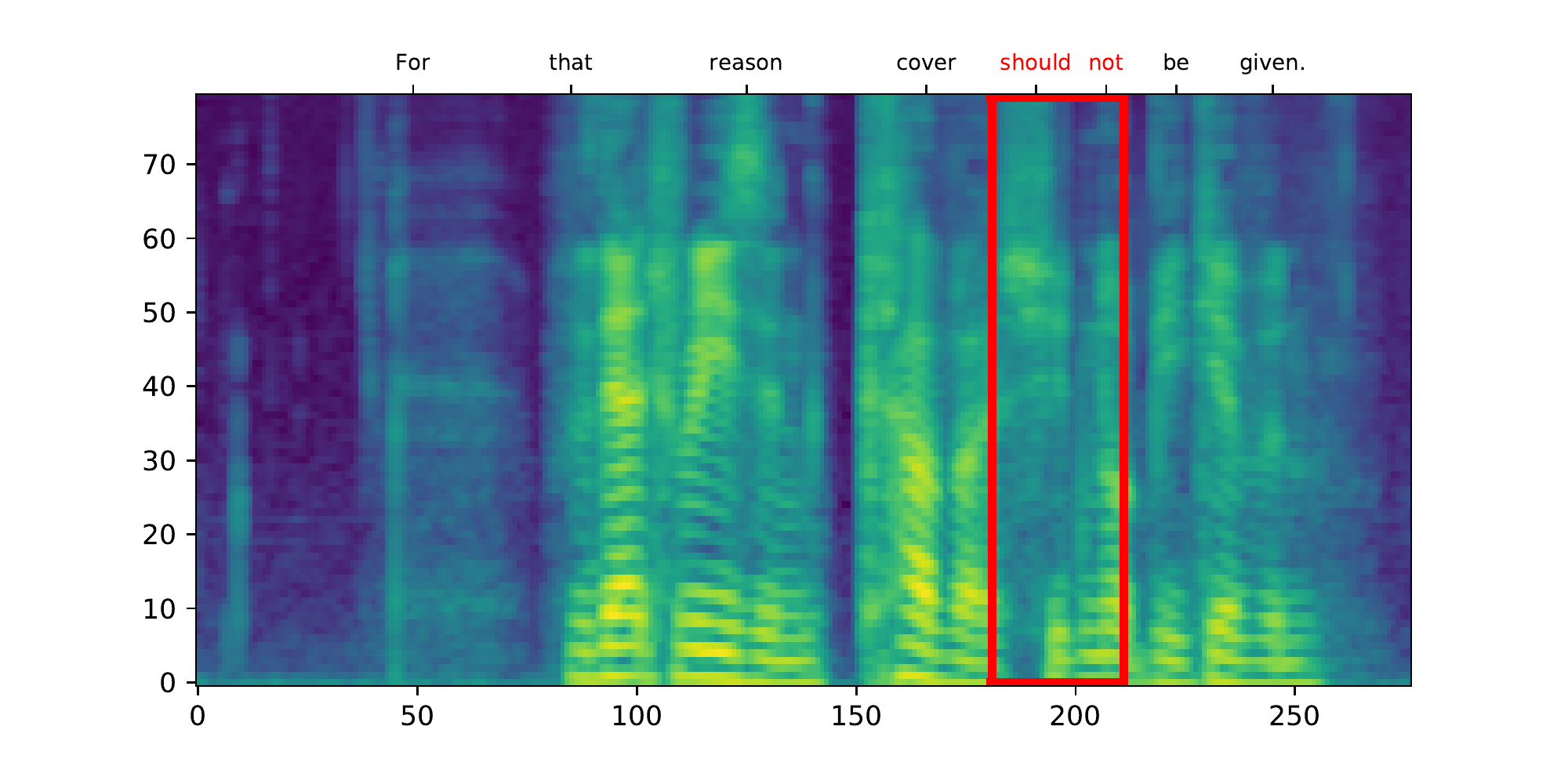}
    }
}
\vspace{-.1cm}
\end{center}
\end{minipage}
\\
\begin{minipage}[t]{.9 \linewidth}
\begin{center}
\vspace{-.5cm}
\subfigure[
Generated modified spectrogram by \aaat.
]{
\makebox[\linewidth][c]{
\includegraphics[height=4.5cm]{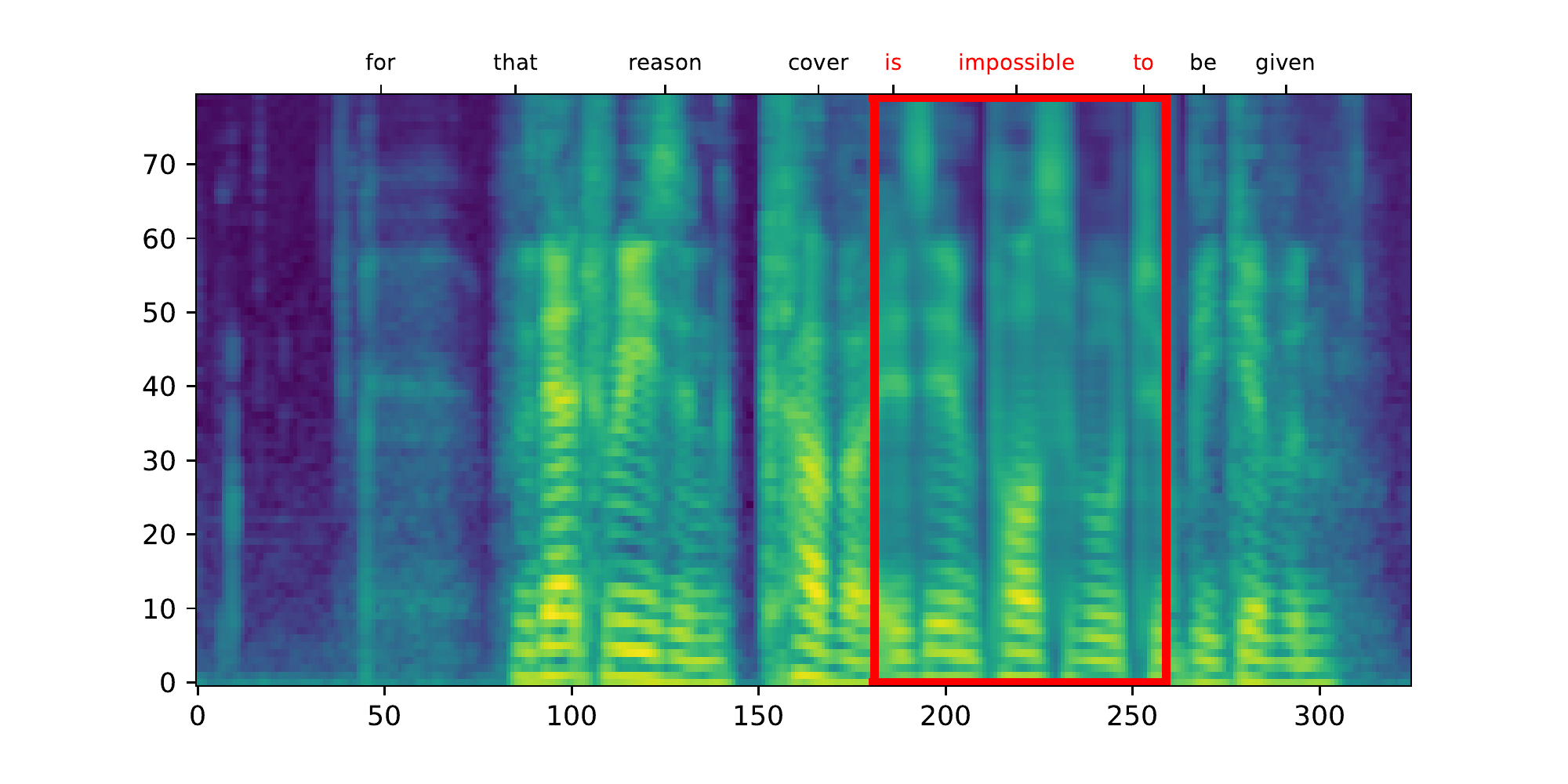}
    }
}
\end{center}
\end{minipage}
\\

\begin{minipage}[t]{.9 \linewidth}
\begin{center}
\vspace{-.5cm}
\subfigure[
Generated modified spectrogram by \citet{tan2021editspeech}.
]{
\makebox[\linewidth][c]{
\includegraphics[height=4.5cm]{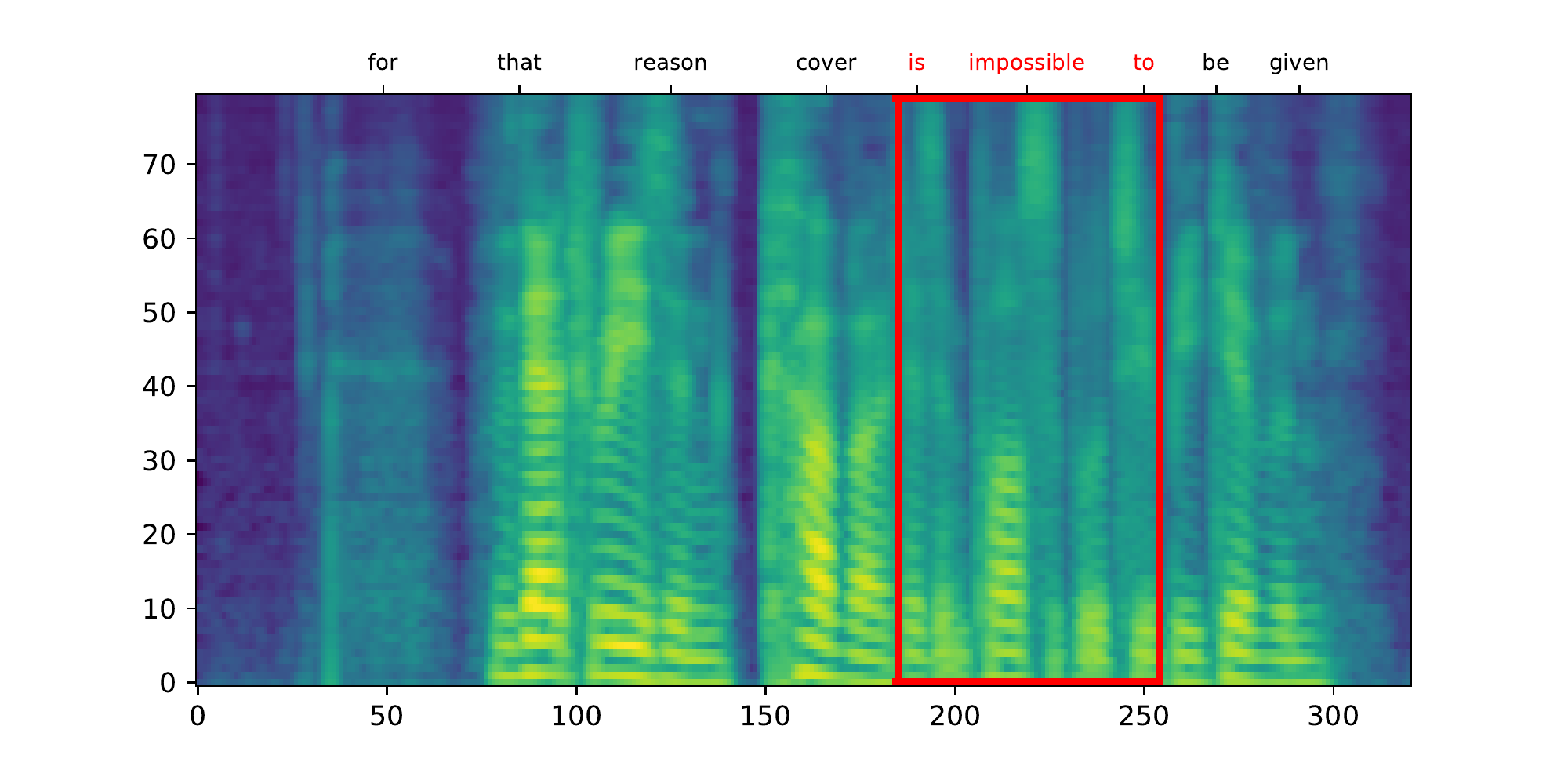}
    }
}
\end{center}
\end{minipage}
\end{tabular}
\caption{
    An speech editing example from VCTK.
    Original text: For that reason cover  \textcolor{red}{should not}  be given.
    Modified text: For that reason cover \textcolor{red}{is impossible to} be given.
}
\label{fig: appendix_spec1}
\end{figure}

\begin{figure}[hb!]
\centering
\begin{tabular}{c}

\begin{minipage}[t]{.9 \linewidth}
\begin{center}
\subfigure[
Groundtruth spectrogram from VCTK.
]{
    \makebox[\linewidth][c]{
\includegraphics[height=4.5cm]{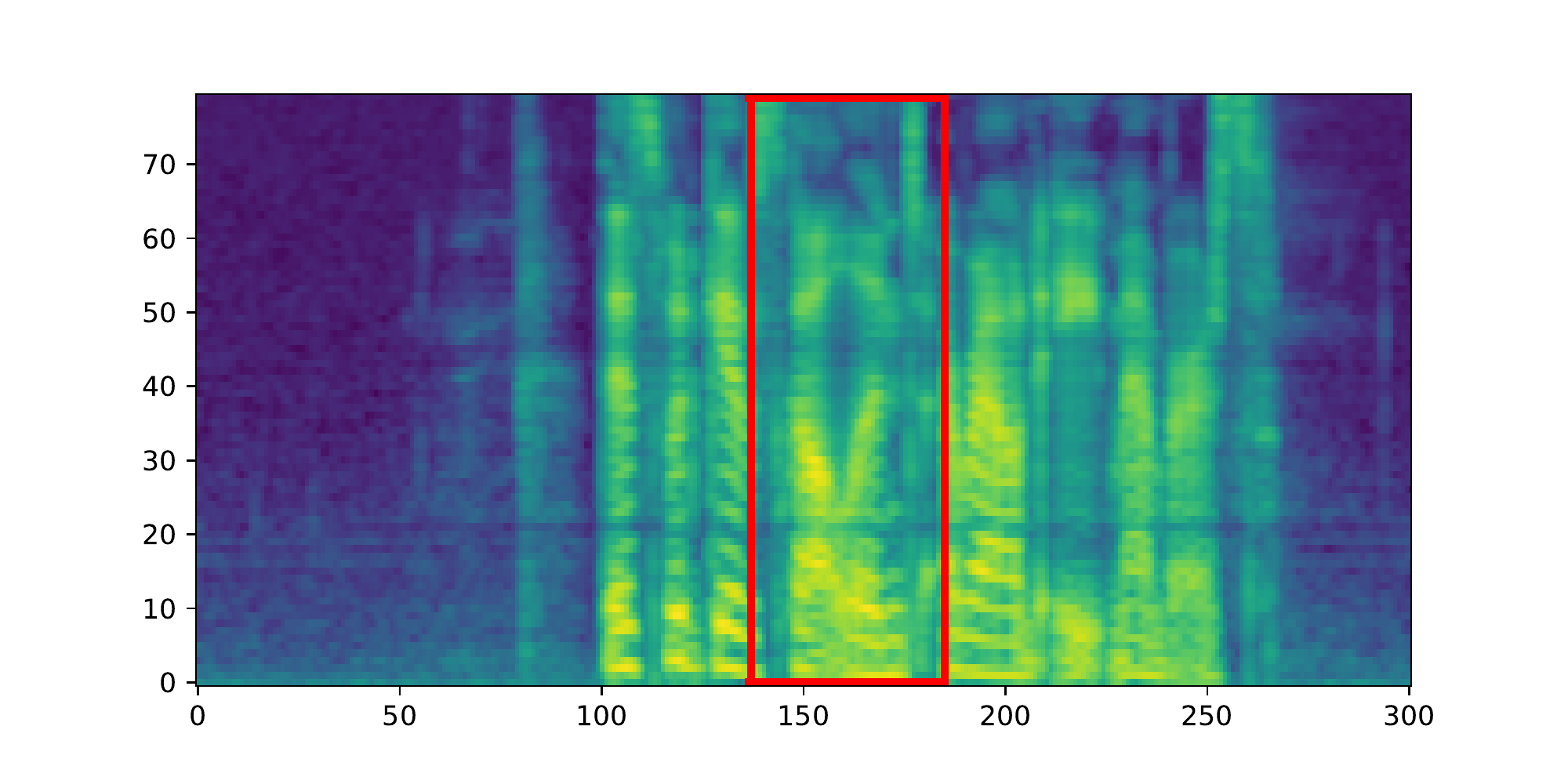}
    }
}
\end{center}
\end{minipage}
\\

\begin{minipage}[t]{.9 \linewidth}
\begin{center}
\subfigure[
Reconstructed spectrogram by \aaat.
]{
\makebox[\linewidth][c]{
\includegraphics[height=4.5cm]{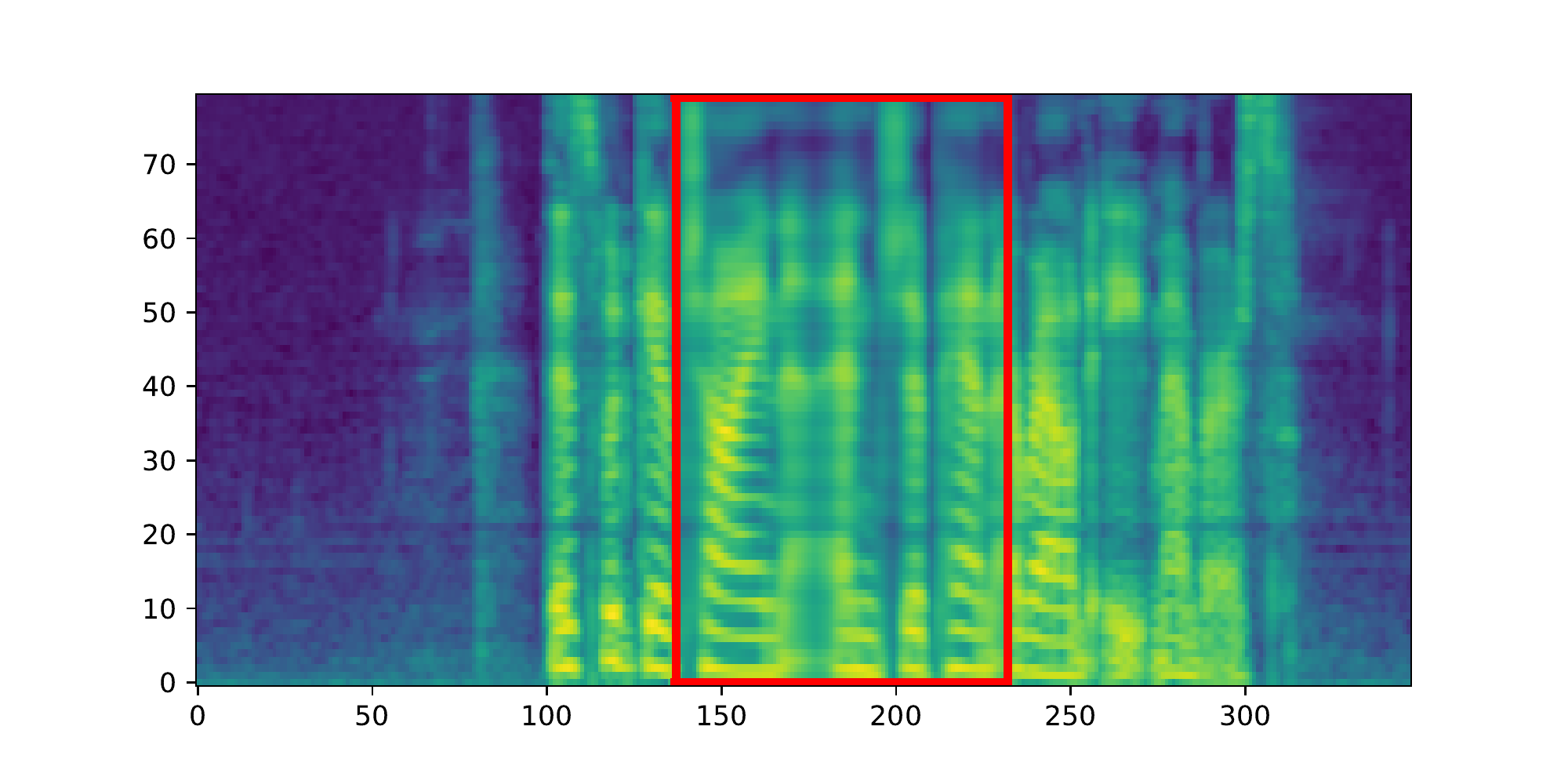}
    }
}
\end{center}
\end{minipage}
\\

\begin{minipage}[t]{.9 \linewidth}
\begin{center}
\subfigure[
Reconstructed spectrogram by \citet{tan2021editspeech}.
]{
\makebox[\linewidth][c]{
\includegraphics[height=4.5cm]{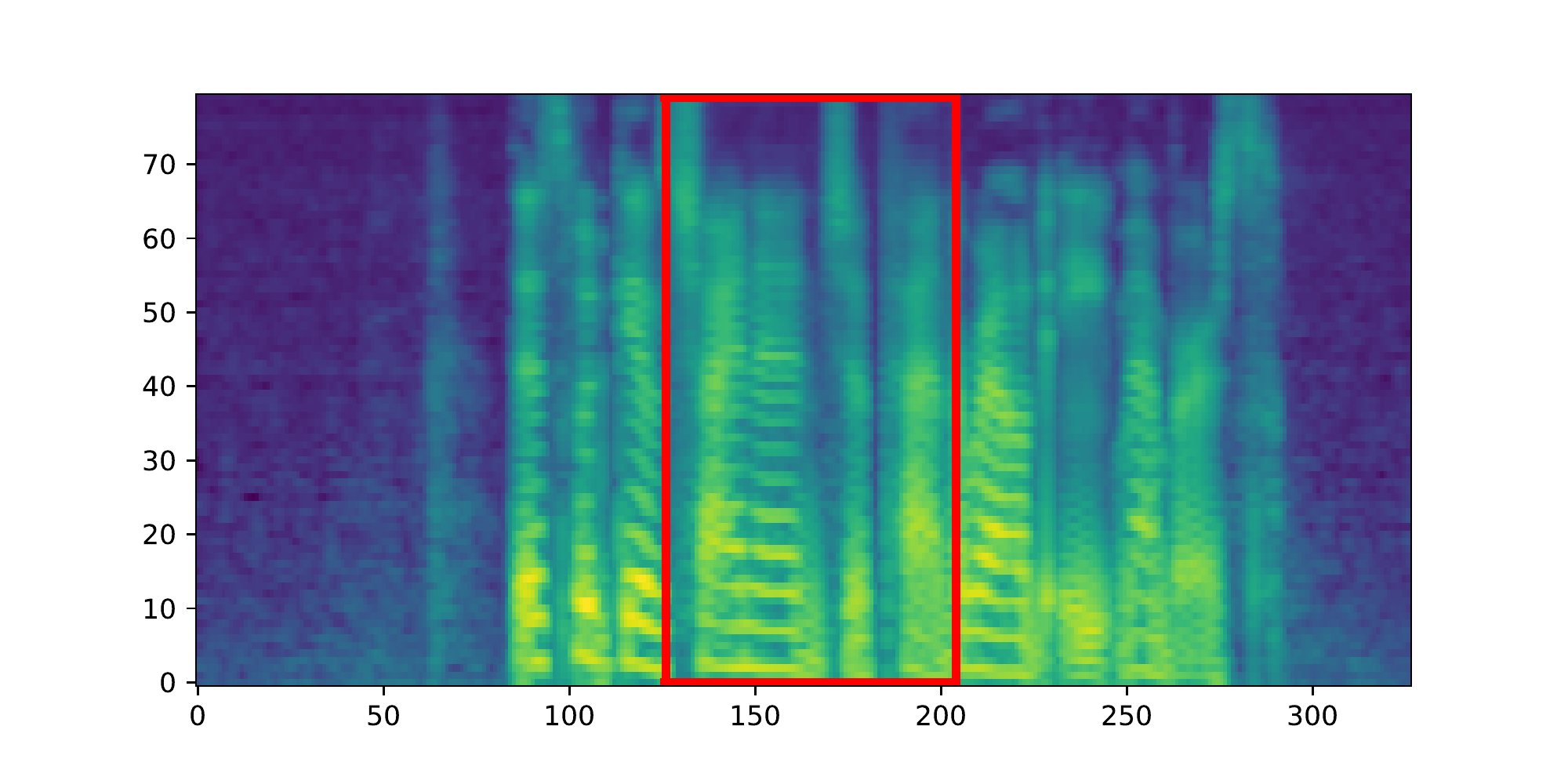}
    }
}
\end{center}
\end{minipage}
\end{tabular}
\caption{
Original text: This would give \textcolor{red}{Scotland} around eight members.
Modified text: This would give \textcolor{red}{China and Japan} around eight members.
}
\label{fig: appendix_spec2}
\end{figure}

\clearpage
\newpage

\section{Pretraining for Multi-speaker TTS}
\label{appendix: tts finetuning}
\begin{table}[ht!]
	\centering
\resizebox{0.7\columnwidth}{!}{
		\begin{tabular}{l l |c}
\toprule
\textbf{Method} &  \textbf{Pretrain Data}  & \textbf{MOS $\uparrow$} \\ 
	\midrule
\multicolumn{2}{l|}{\textit{Groundtruth}}   & 4.07 $\pm$ 0.07 \\ 
FastSpeech 2 & $\mathtt{none}$            &  3.63 $\pm$ 0.07 
 \\ \midrule 
 \aaat & LibriTTS          &   3.72 $\pm$ 0.07 
\\
\aaat & ASR + Speech &    \textbf{3.77} $\pm$ 0.07 
 \\
\bottomrule
\end{tabular}
}
\caption{MOS evaluation  for multi-speaker speech synthesis.}\label{table:libritts_mos}
\end{table}

\begin{figure}[!ht]
  \centering
\subfigure[Training Loss]{
\includegraphics[width=0.23\textwidth
]{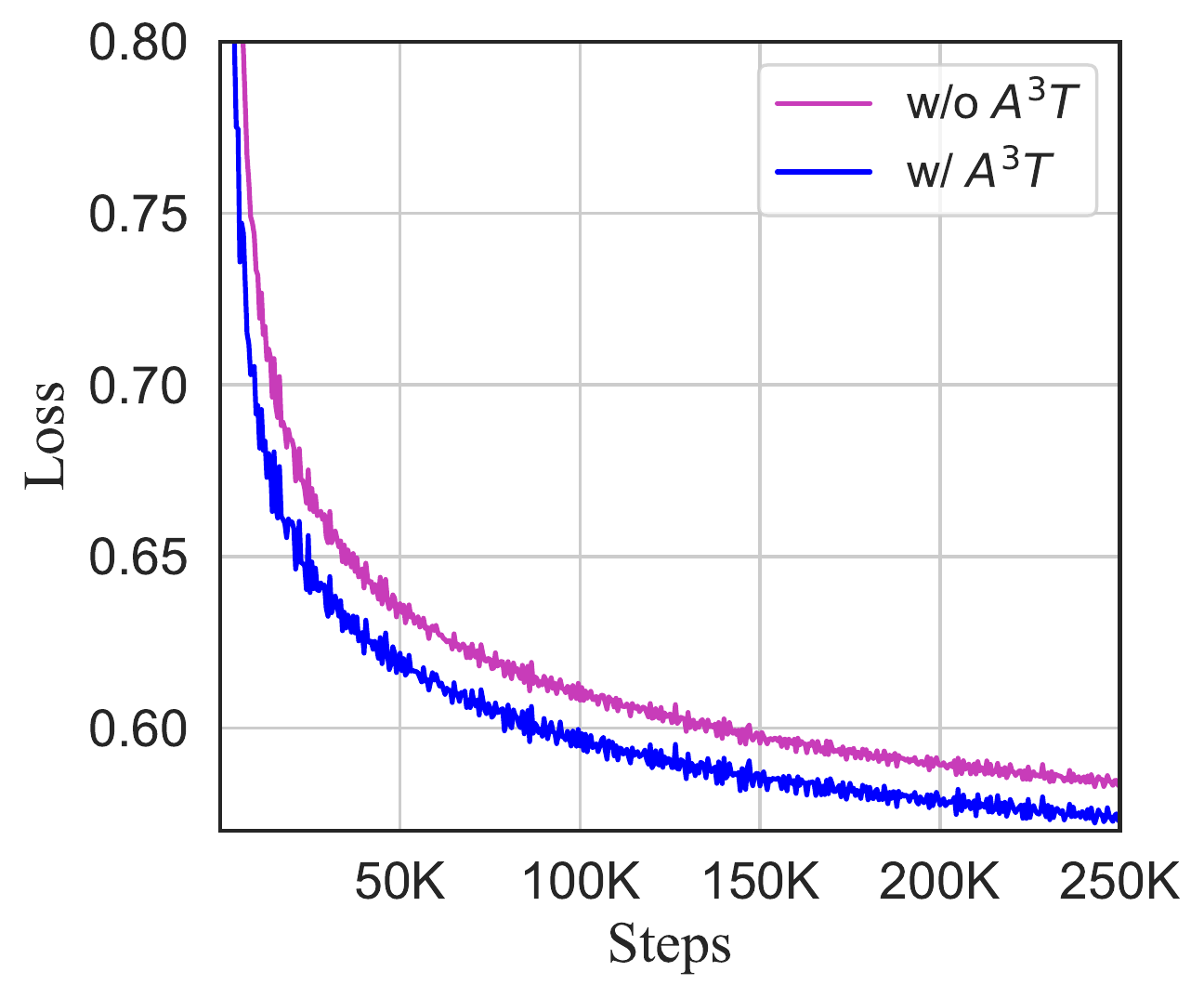}}
\subfigure[Validation Loss]{
\includegraphics[width=0.23\textwidth
]{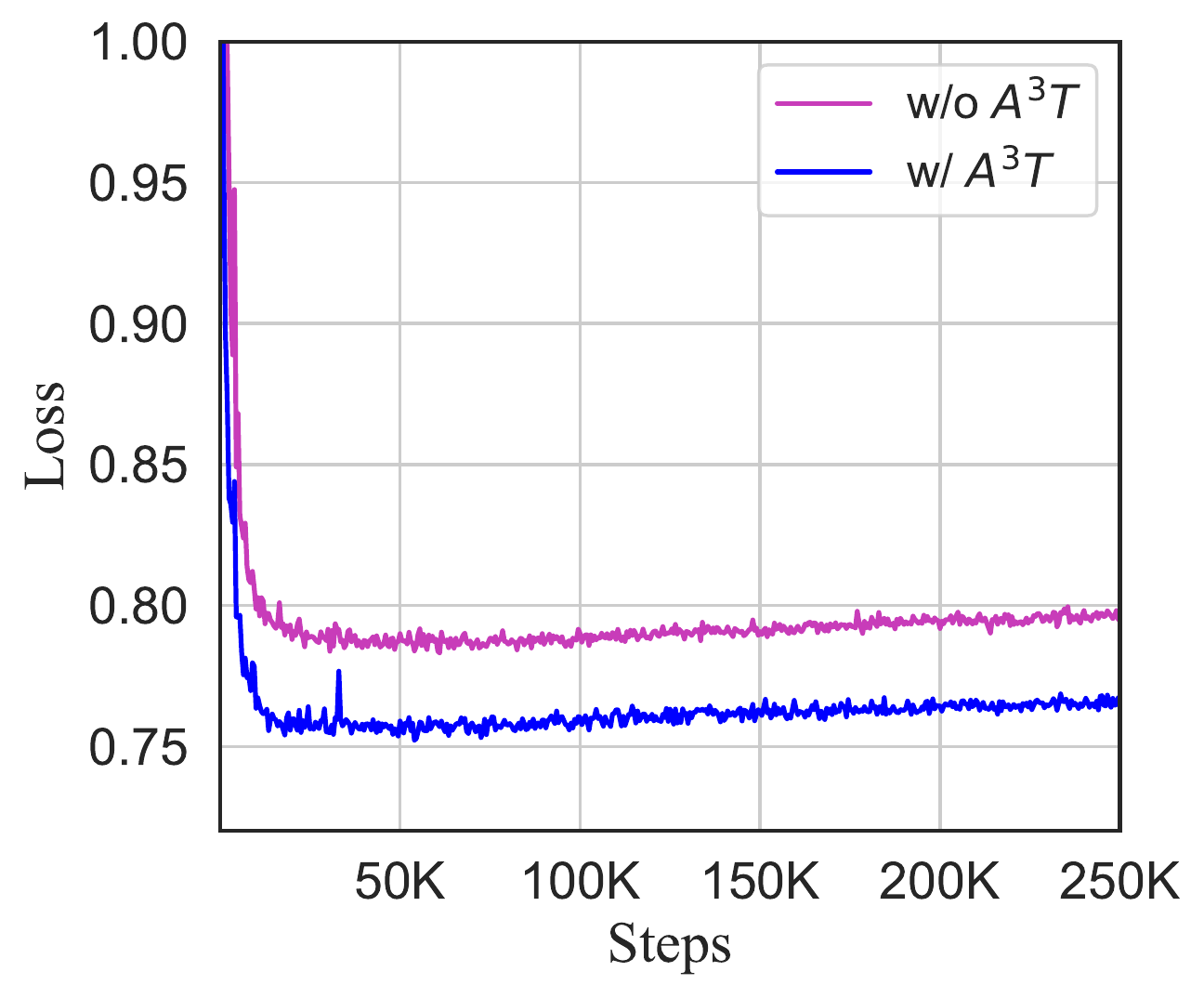}}
  \caption{Training and validation loss using LibriTTS between TTS models initialized with and without \aaat.}\label{fig: loss_libritts}
\end{figure}

We conduct finetuning experiments with a large multi-speaker TTS dataset LibriTTS 
but split the validation and test set with only the new speakers.
We test on 50 test cases with 15 human annotators for each case.
In this setting, we find the FastSpeech 2 fails to generate high-quality audio 
for these new speakers, even equipped with X-Vector~\cite{snyder2018x} to generate speaker embeddings for new speakers. 
After initializing the FastSpeech 2 model with our LibriTTS pretrained \aaat ,
the generated audio can be improved significantly. 
Results are shown in Tab.~\ref{table:libritts_mos}.
We also plot the validation loss and training loss during the training of TTS models with and without \aaat in Fig.~\ref{fig: loss_libritts}.
We can see that both the training and validation loss is improved with the initialization from the \aaat model, which demonstrates the effectiveness of our method. 
Finally, we also observe the improvement from the external data~(LibriSpeech~\cite{panayotov2015librispeech} and LibriLight~\cite{librilight}) pretraining for the \aaat model, which achieves 3.77 MOS scores in Tab.~\ref{table:libritts_mos}.
It should be noted that when training with speech only data LibriLight, our model is similar to MAM~\cite{chen2020mam} and the alignment embedding are discarded.

\end{document}